\documentclass[fleqn,usenatbib]{mnras}

\usepackage{newtxtext,newtxmath}

\usepackage[T1]{fontenc}

\DeclareRobustCommand{\VAN}[3]{#2}
\let\VANthebibliography\thebibliography
\def\thebibliography{\DeclareRobustCommand{\VAN}[3]{##3}\VANthebibliography}

\usepackage{graphicx}	
\usepackage{amsmath,bm}	
\usepackage{caption}
\usepackage{subcaption}

\newcommand{\kmsmpc}{km s$^{-1}$ Mpc$^{-1}$}
\defcitealias{Grayling24}{G24}

\title{BayeSN-TD: Time Delay and $H_0$ Estimation for Lensed SN H0pe}

\author[M. Grayling et al.]{
M. Grayling$^1$\thanks{Email: mg2102@cam.ac.uk}, 
S. Thorp$^{1,2}$, 
K. S. Mandel$^{1,3}$, 
M. Pascale$^4$,
J. D. R. Pierel$^5$,
E. E. Hayes$^{1}$, 
 \newauthor \
C. Larison$^5$,
A. Agrawal$^{6, 7}$, 
G. Narayan$^{6,7}$
\\
$^1$ Institute of Astronomy and Kavli Institute for Cosmology, Madingley Road, Cambridge CB3 0HA, UK \\
$^2$ The Oskar Klein Centre, Department of Physics, Stockholm University, AlbaNova University Centre, SE 106 91 Stockholm, Sweden \\
$^3$ Statistical Laboratory, DPMMS, University of Cambridge, Wilberforce Road, Cambridge, CB3 0WB, UK\\
$^4$ Department of Physics \& Astronomy, University of California, Los Angeles, 430 Portola Plaza, Los Angeles, CA 90095, USA \\
$^5$ Space Telescope Science Institute, Baltimore, MD 21218, USA \\
$^6$ Department of Astronomy, University of Illinois Urbana-Champaign, 1002 West Green Street, Urbana, IL 61801, USA \\
$^7$ NSF-Simons AI for the Sky (SkAI) Institute, 875 N. Michigan Ave., Suite 3500, Chicago, IL 60611, USA \\
}

\date{Accepted XXX. Received YYY; in original form ZZZ}

\pubyear{2025}

\begin{document}
\label{firstpage}
\pagerange{\pageref{firstpage}--\pageref{lastpage}}
\maketitle

\begin{abstract}
We present BayeSN-TD, an enhanced implementation of the probabilistic type Ia supernova (SN Ia) BayeSN SED model, designed for fitting multiply-imaged, gravitationally lensed type Ia supernovae (glSNe Ia). BayeSN-TD fits for magnifications and time-delays across multiple images while marginalising over an achromatic, Gaussian process-based treatment of microlensing, to allow for time-dependent deviations from a typical SN Ia SED caused by gravitational lensing by stars in the lensing system. BayeSN-TD is able to robustly infer time delays and produce well-calibrated uncertainties, even when applied to simulations based on a different SED model and incorporating chromatic microlensing, strongly validating its suitability for time-delay cosmography. We then apply BayeSN-TD to publicly available photometry of the glSN Ia SN H0pe, inferring time delays between images BA and BC of $\Delta T_{BA}=121.9^{+9.5}_{-7.5}$ days and $\Delta T_{BC}=63.2^{+3.2}_{-3.3}$ days along with absolute magnifications $\beta$ for each image, $\beta_A = 2.38^{+0.72}_{-0.54}$, $\beta_B=5.27^{+1.25}_{-1.02}$ and $\beta_C=3.93^{+1.00}_{-0.75}$. Combining our constraints on time-delays and magnifications with existing lens models of this system, we infer $H_0=69.3^{+12.6}_{-7.8}$ \kmsmpc, consistent with previous analysis of this system; incorporating additional constraints based on spectroscopy yields $H_0=66.8^{+13.4}_{-5.4}$ \kmsmpc. While this is not yet precise enough to draw a meaningful conclusion with regard to the `Hubble tension', upcoming analysis of SN H0pe with more accurate photometry enabled by template images, and other glSNe, will provide stronger constraints on $H_0$; BayeSN-TD will be a valuable tool for these analyses. The BayeSN-TD code is available at \url{https://github.com/bayesn/bayesn-td}.

\end{abstract}

\begin{keywords}
gravitational lensing: strong,
supernovae: individual: SN H0pe,
methods: statistical
\end{keywords}



\section{Introduction}
\label{sec:Intro}

Strong gravitational lensing is a phenomenon whereby a massive system, such as a galaxy or galaxy cluster, lies along the line-of-sight between an observer and an astronomical source. The gravitational impact of the lensing system magnifies the background source and causes multiple images of it to appear. In the event of a time-varying event such as a quasar or supernova, multiple images of the system will appear with observable time offsets caused by differences in geometry and gravitational potential between the paths travelled through the lensing system to reach the observer. It was first proposed by \citet{Refsdal64} that the Hubble constant $H_0$ could be measured by combining the time delay between multiple images of a gravitationally lensed supernova (glSN) with a model of the mass distribution of the lensing system. The potential of glSNe to provide an independent measurement of $H_0$ is invaluable given the ongoing tension between early-Universe measurements from the cosmic microwave background \citep[CMB;][]{Planck20} and local measurements based on the distance ladder using SNe Ia \citep[e.g.][]{Reiss22, Li25} (although this is not reflected across all distance ladder measurements, see e.g. \citealp{Freedman24}; for a full review including other distance indicators please see \citealp{Valentino21}). In this work we present BayeSN-TD, an enhanced version of the probabilistic type Ia supernova (SN Ia) spectral energy distribution (SED) model, BayeSN \citep{M20, T21, Grayling24}, adapted for fitting light curves of glSNe, and validate the performance of this model through application to simulations. We then apply this model to obtain constraints on the time delays and magnifications of SN H0pe using photometry from \citet{Pierel24}, along with corresponding constraints on $H_0$ by combining these with the lens models of SN H0pe presented in \citet{Pascale25}.

Any gravitationally-lensed time-varying event could in principle be used to estimate a time delay and consequently $H_0$; gravitationally-lensed quasars have previously been used to infer $H_0$ in a number of different studies \citep[e.g.][]{Keeton97, Wong20, Birrer20, TDCosmo25} - see \citet{Birrer24} for a recent review. However, glSNe have several advantages over quasars for these analyses. One reason is that SNe fade, allowing for isolated analysis of the lens and host and more accurate photometry using a template, compared with the blended light from a quasar, lens and host \citep{Ding21}. SNe also have simpler light curves with variability over weeks to months, compared with the longer-term stochastic variation from quasars - this simplifies time-delay estimation and means that shorter observing campaigns are required for these measurements. The more compact size of the source for glSNe compared with quasars also reduces the impact of microlensing on time-delay estimates \citep{Tie18, Bonvin19}.

For the special case of a gravitationally-lensed type Ia supernova (glSN Ia), we can additionally use the standardisable nature of these events to constrain the absolute magnification of each image, providing additional constraint on the lens model and limiting the uncertainty caused by the mass sheet degeneracy \citep[e.g.][]{Falco85, Kolatt98, Holz01, Oguri03, Nordin14}.

However, some of the advantages to using glSNe to constrain $H_0$ also lead to inherent challenges. The fact that these events fade on relatively fast timescales -- in contrast to lensed quasars -- also makes them more difficult to discover, especially considering their rarity. To date, only a small number of glSNe have been observed. The first resolved, multiply-imaged glSN was SN Refsdal, a peculiar type II SN. Analysis of this object led to estimates of $H_0=64.8^{+4.4}_{-4.3}$~\kmsmpc or $H_0=66.6^{+4.1}_{-3.3}$~\kmsmpc depending on the lens model weights \citep{Kelly23a,Kelly23b}. A number of other glSNe have been observed which were not suitable for $H_0$ inference; for example iPTF16geu \citep{Goobar17} and SN Zwicky \citep{Goobar23, Pierel23, Larison25} were spectroscopically-confirmed glSNe with very short time delays of $\sim0.25-2.0$ days which prevented $H_0$ estimates of reasonable precision \citep{Dhawan20, Pierel23}.

The first glSN Ia with time delays long enough to enable a competitive $H_0$ analysis was SN H0pe, discovered in March 2023 \citep{Frye23} by the `Prime Extragalactic Areas for Reionization and Lensing Science' \citep[PEARLS; PID 1176,][]{Windhorst23} James Webb Space Telescope (JWST) programme. SN H0pe was followed up in a DDT programme (PID 4446, PI: B. Frye) for two additional epochs of NIRCam photometry and one epoch of NIRSpec spectroscopy. Template photometry of the field was obtained in a recent programme (PID 4744; PIs: B. Frye and J. Pierel). Photometric and spectroscopic analysis, the complete set of lensing evidence, and the first lens model are presented in \citet{Frye24}. \citet{Pierel24} presented time-delay and magnification estimates using photometric observations, \citet{Chen24} presented time-delay analysis using spectroscopic observations and \citet{Pascale25} used these time-delay estimates to infer $H_0$. This analysis led to an inferred value of $H_0=75.4^{+8.1}_{-5.5}$ \kmsmpc when leveraging absolute magnitude information about SNe Ia and $H_0=71.8^{+9.8}_{-7.6}$ \kmsmpc without using this information. Since the discovery of SN H0pe, another glSN Ia with time-delays suitable for $H_0$ inference has been identified; SN Encore \citep{Pierel24b}, which notably occurred in the same galaxy as SN Requiem \citep{Rodney21}. \citet{Pierel25} and \citet{Suyu25} respectively present time-delays and lens models for SN Encore which yield an estimate of $H_0=66.9^{+11.2}_{-8.1}$ \kmsmpc.

Given the serendipitous discovery of SN H0pe, photometric data for two of the three images is only available significantly after peak, extending out to a phase of $\sim$ +60 days. This necessitates the use of a SN SED model which extends both to near infrared (NIR) wavelengths and late-time phases, making BayeSN the only viable current SN Ia SED model for this analysis. \citet{Pierel24} applied an extended version of the BayeSN model presented in \citet{W22} which covered phases out to 50 rest-frame days after peak and used linear extrapolation beyond that range. However, for analyses of such events it is desirable to extend the coverage of the model to later phases.

Another significant challenge when fitting light curves of glSNe with typical SN models is the presence of microlensing -- lensing caused by small perturbers in the lens plane such as stars -- which can have a time-varying impact across the observed SN light curve. This can significantly affect inferred time delays from glSNe \citep[e.g.][]{Dobler06, Goldstein18, Pierel19} and must be accounted for in such analyses \citep[see e.g. Fig 1 of][]{Hayes24}. Microlensing can cause otherwise typical SNe to appear significantly different with deviations from typical SN SEDs. It is typical to fit SN Ia light curves using empirical SED models such as BayeSN \citep{T21, M20,Grayling24} or SALT \citep{Guy07, Kenworthy21}, trained on populations of non-lensed SNe Ia. Naively applying these models to glSNe without mitigating the impact of microlensing could potentially lead to biased time-delay estimates and underestimated uncertainties. 

Previous studies have found that, for SNe Ia, microlensing is effectively achromatic for approximately the first 3 weeks after explosion \citep{FoxleyMarrable18, Goldstein18, Huber19, Huber21} - the impact on a SN light curve is the same in each band. Supernova Time Delays (\textsc{SNTD}), presented in \citet{Pierel19}, mitigates the impact of microlensing by applying SN models to measured colours rather than measured photometry, as colours are insensitive to achromatic microlensing \citep{Goldstein18}. The impact of microlensing is also effectively achromatic during the plateau phase of SNe IIP \citep{Bayer21}, and \citet{Grupa25} analysed colour curves for time-delay inference of simulated SNe IIP light curves. \citet{Huber22, Huber24} followed a different approach, training machine learning methods for time-delay estimation using physical simulations of SN observables which incorporate the effect of chromatic microlensing using intensity profiles of theoretical models. \citet{Hayes24} applied a template-independent method for time-delay inference based on Gaussian processes (GPs), using an analytic, achromatic treatment of microlensing. This has since been enhanced to incorporate templates along with a chromatic, GP-based treatment of microlensing \citep{Hayes25b}.

In this work, we present BayeSN-TD, a method for fitting glSNe Ia which combines the probabilistic SN Ia SED model BayeSN with a GP treatment of microlensing to allow for deviations from the SED model. This approach models the underlying SN Ia light curve along with variations between each image; this enables joint inference of typical SN Ia standardisation parameters such as light curve shape, time delays between each image and the impact of microlensing on each image. Performing these fits as a single joint inference ensures that the impact of microlensing on the time delay and overall SN light curve is incorporated robustly and forms part of the statistical error budget when estimating $H_0$. For this work we assume an achromatic treatment of microlensing. We validate the performance of BayeSN-TD through application to simulations of glSNe Ia which incorporate the effect of microlensing, demonstrating that this model is able to robustly infer time delays with well-calibrated uncertainties. We also introduce a new extended version of the BayeSN model with coverage to a phase of +85 days for use in cases where only late-time photometry is available. Having validated the performance of BayeSN-TD through simulations, we apply this new model to the photometry of SN H0pe presented in \citet{Pierel24} to obtain estimates of time delays and magnifications, and combine these with the lens models presented in \citet{Pascale25} to obtain an associated constraint on $H_0$.

The structure of this paper is as follows. In Section \ref{sec:method} we detail the BayeSN-TD model and its implementation, as well as the phase-extended version of the BayeSN model used for this work. We then validate the performance of our model on a variety of simulated glSNe in Section \ref{sec:sims}, before applying it to photometry of SN H0pe from \citet{Pierel24} in Section \ref{sec:h0pe} to obtain time-delay and magnifications estimates. We then present corresponding constraints on $H_0$ from our time-delays and magnifications in Section \ref{H0}. Finally, we conclude in Section \ref{sec:conclusions}.

\section{Method}
\label{sec:method}

We begin by giving an overview of the BayeSN model which forms the basis for this work, and detailing how we have extended it with BayeSN-TD for application to strongly lensed SNe Ia. The full description of the BayeSN SED model is presented in \citet{M20}\footnote{Building upon earlier hierarchical Bayesian multi-passband SN Ia light curve models of \citet{Mandel09, Mandel11}}, with further discussions in \citet{T21, TM22, W22, Thorp24, Grayling24, Uzsoy24, Hayes25, GraylingPopovic25}.

\subsection{The BayeSN Model}

BayeSN is a probabilistic SED model for SNe Ia, with the full time- and wavelength-varying SED given by:
\begin{equation}
\label{bayesn_equation}
\begin{aligned}
    -2.5\log_{10}[S_s(t,\lambda_r)/S_0(t,\lambda_r)] = M_0 + W_0(t,\lambda_r) \ + \ \\ \delta M^s+\theta^s_1W_1(t,\lambda_r) + \epsilon^s(t,\lambda_r)+A^s_V\xi\big(\lambda_r;R^{(s)}_V\big)
\end{aligned}
\end{equation}
where $t$ signifies the phase relative to B-band maximum, and $\lambda_r$ denotes the rest-frame wavelength. BayeSN uses the optical-NIR SN Ia SED template from \citet{Hsiao07} as a zeroth-order template, with an arbitrary scaling factor $M_0$ of -19.5\footnote{Note that $M_0$ does not define the absolute magnitude scale of SNe Ia, this is arbitrarily fixed to -19.5 with $W_0$ then defining the mean intrinsic SED for the population.}. Latent variables, with distinct values for each SN, are denoted by superscript $s$, while all other parameters are global hyperparameters that are shared across the population. The individual components constituting the model are detailed below:

\begin{itemize}
    \item The function $W_0(t, \lambda_r)$ warps and normalises the zeroth-order SED template, which establishes a mean intrinsic SED for the SN Ia population. $W_1(t, \lambda_r)$ is a functional principal component (FPC) designed to capture the primary mode of intrinsic SED variation across the population of SNe Ia. These components are both implemented as cubic spline surfaces.
    \item For each SN, the coefficient $\theta_1^s$ quantifies the impact of the $W_1$ FPC. This coefficient is defined with a Normal prior distribution such that $\theta_1^s \sim N(0, 1)$. When combined, $W_1(t, \lambda_r)$ and $\theta_1^s$ effectively model the `broader-brighter' relationship inherent to SNe Ia, where intrinsically brighter light curves are observed to evolve over more extended timescales around their peak \citep{Phillips93}.
    \item $\delta M^s$ is an achromatic, time-independent magnitude offset for each SN, drawn from a normal distribution with $\delta M^s \sim N(0, \sigma_0^2)$. The hyperparameter $\sigma_0$, which defines the intrinsic achromatic scatter across the population, is inferred during the model's training phase.
    \item The term $\epsilon^s(t,\lambda_r)$ is a time- and wavelength-dependent function that describes residual intrinsic colour variations within the SED that are not accounted for by the $\theta_1^sW_1(t, \lambda_r)$ component. This parameter is represented by a cubic spline function over time and wavelength, which is defined by a matrix of knots, $\mathbf{E}^s$. These knots are drawn from a multivariate Gaussian distribution, $\mathbf{e}^s \sim N(0, \mathbf{\Sigma}_\epsilon)$, where $\mathbf{e}^s$ is the vectorised version of the $\mathbf{E}^s$ matrix. The covariance matrix $\mathbf{\Sigma}_\epsilon$ functions as a model hyperparameter that characterises the distribution of this residual scatter across the SN Ia population, and is inferred during training.
    \item The host galaxy extinction law for each supernova is described by $A_V^s$ and $R_V^{(s)}$. $A_V^s$ represents the total V-band extinction amount, while $R_V^{(s)}$ describes the slope of the \citet{Fitzpatrick99} dust extinction law assumed by the model. $R_V^{(s)}$ can be treated as either a shared parameter for the whole population or as a latent parameter for each supernova drawn from a distribution. For $A_V^s$, an exponential prior is assumed, governed by a scale parameter $\tau_A$, such that $A_V^s \sim \text{Exponential}(\tau_A)$.
\end{itemize}

An advantage of BayeSN is that it models the physically-distinct effects of intrinsic variations and host-galaxy dust on the supernova SED when fitting SN Ia light curves \citep[e.g.][]{Mandel17}.

The rest-frame, host galaxy dust-extinguished SED model $S_s(t, \lambda_r)$ is then scaled based on distance modulus $\mu^s$, redshifted and corrected for Milky Way dust extinction assuming $R_V=3.1$, using a \citet{Fitzpatrick99} dust law with dust maps from \citet{Schlafly11}. Model photometry can be derived by integrating this SED through photometric filters, which can then be compared with observed photometry to compute a likelihood. BayeSN training involves jointly inferring all global and latent parameters across a population of SNe Ia (for a complete discussion on model training, see \citealp{M20, T21}). We marginalise over all latent parameters and obtain estimates of global parameters from the posterior distributions. Once trained, BayeSN can be applied to fit light curves of individual SNe by inferring posteriors of the latent parameters for each SN conditional on the fixed population-level parameters inferred during model training.

\subsection{Improving Phase Coverage of BayeSN}
\label{extended_bayesn}

As discussed in \citet{Pierel24}, one challenge faced by analysis of SN H0pe and other strongly lensed SNe discovered by JWST is a lack of coverage of SN Ia SED models at later phases, especially in NIR wavelengths. \citet{Pierel24} applied a phase-extended version of the BayeSN model presented in \citet{W22}, which was defined up to 50 rest-frame days after peak, and utilised linear extrapolation beyond this phase. As part of this work, we train a new BayeSN model with later phase coverage extending out to 85 rest-frame days after peak. We apply this model within BayeSN-TD for time delay estimation of SN H0pe in this work and make it publicly available for future analyses. This phase-extended BayeSN model has been used as the basis for fitting another glSN Ia, SN Encore, as presented in \citet{Pierel25}.

Previous trained BayeSN models were presented in \citet{M20}, \citet{T21} and \citet{W22}. \citet{M20} trained on a compilation of local SNe Ia presented in \citet{Avelino19}, while \citet{T21} trained on a sample of SNe Ia from Foundation DR1 \citet{Foley18} and \citet{W22} trained on the combination of those two datasets. In addition, \citet{TM22} applied the model presented in \citet{M20} to a sample of SNe Ia exclusively from CSP-I \citep{Krisciunas17}, selecting a sample from that presented in \citet{Uddin20}. Within this work, we train on a combination of all SNe across these analyses, yielding a total training set of 278 SNe Ia. Of these, those from Foundation have only optical photometry while the rest also have NIR (YJH bands).

In terms of technical implementation, this model is very similar to that of \citet{W22} except for the addition of extra spline knots at later phases of +55, +70 and +85 days when defining $W_0(t,\lambda_r)$, $W_1(t,\lambda_r)$ and $\mathbf{\Sigma}_\epsilon$ to allow the later phase coverage. One other difference is that we include U-band data in the training set, unlike \citet{M20} and \citet{W22} but similarly to the phase-extended model applied in \citet{Pierel24}. We make this choice to include F090W data within our analysis, as F090W data for SN H0pe covers rest-frame U-band. Full details around training the BayeSN model are presented in \citet{M20}, while the code used for training the model is described in \citet{Grayling24}.

This new, phase-extended BayeSN model can be found at \url{https://github.com/bayesn/bayesn-model-files/tree/main}, and is incorporated within the public BayeSN code available here: \url{https://github.com/bayesn/bayesn}. For more discussion of this new BayeSN model, please see Appendix \ref{appendix:extended_bayesn}.

\subsection{Fitting Multiple Images Using BayeSN-TD}

When applying a trained BayeSN model to fit a single SN Ia light curve, a number of different latent SN parameters are inferred: the `shape' parameter $\theta_1^s$, $V$-band host galaxy dust extinction $A_V^s$ (and, optionally, the slope of the dust extinction law $R_V^s$), the distance modulus $\mu^s$, the residual intrinsic colour surface $\epsilon^s(t,\lambda_r)$ and finally the time of B-band maximum $t^s_\text{max}$. When fitting a multiply-imaged type Ia supernova, many of these parameters are treated as being shared across each image i.e. we are seeing multiple images of the same intrinsic SN light curve. However, separate parameters are included for different images of the same SN to account for time delays and magnification. A full description of the parameters which are shared between images and those which vary between images is given below. Please note that the index $s$ denotes parameters shared across all images of a SN $s$, while the index $i$ denotes parameters which differ between separate images of the same SN.

\begin{itemize}
    \item \textbf{Parameters shared between images}
    \begin{itemize}
        \item Light curve shape $\theta_1^s$
        \item Host galaxy dust extinction parameters $A_V^s$ and $R_V^s$
        \item Residual intrinsic colour $\epsilon^s(t,\lambda_r)$
    \end{itemize}
    \item \textbf{Parameters varying between images}
    \begin{itemize}
        \item Time of B-band maximum $t_\text{max}^{si}$
        \item Distance modulus $\mu^{si}$ is treated separately for each image to allow for  differences in magnification.
        \item BayeSN-TD incorporates an analytic treatment for the effect of microlensing using a GP, as outlined in Section \ref{gibbs}. Each image of a SN receives its own GP hyperparameters and corresponding microlensing curve.
    \end{itemize}
\end{itemize}

In future, further complexity could be incorporated in the model. For example, we could account for differences in the dust properties in the lens along the line-of-sight to each of the images along with the effect of dust extinction in the host galaxy of the SN; this would require $A_V^s$ and $R_V^s$ parameters for host galaxy extinction for the SN along with separate $A_V^{si}$ and $R_V^{si}$ parameters for each image to capture the separate effect of dust extinction in the lens for each image.

\subsection{Incorporating Microlensing}
\label{gibbs}

The multiple images of a lensed source result from different paths taken by the light from each image through the lens system; light for each image therefore passes through a unique star field, each with its own lensing magnification map. These maps vary on the scale of microarcseconds, comparable to typical physical sizes of the photospheres of SNe. Over time, as the photosphere of a SN expands it passes over an increasing number of microlens caustics. This causes a time-varying magnification which is unique for each image \citep[e.g.][]{Dobler06, Bagherpour06, FoxleyMarrable18}. This effect, microlensing, can have a significant impact on time-delay measurements \citep[e.g.][]{Goldstein18, Pierel19, Hayes24}, and any time-delay analysis of glSNe must consider this impact.

In the standard BayeSN model, variation around the population mean intrinsic SED for SNe Ia is governed by a functional principal component $\theta_1^s$ along with the impact of dust extinction and residual intrinsic scatter $\epsilon^s(t,\lambda_r)$ corresponding to the distribution of intrinsic SN colours. However, when applying BayeSN to strongly lensed supernovae we additionally allow for deviations from the SED model as a result of microlensing.

In this work we opt for a flexible treatment of microlensing using Gaussian processes (GPs). GPs have been used extensively in transient astronomy, for example to model light curves \citep[e.g.][]{Grayling21, Grayling23, Revsbech18, aigrain23}. GPs have also been used extensively for time-delay cosmography, applied to strongly-lensed quasars \citep{Hojjati14, Tak17, Hu20, Meyer23} and SNe \citep{Kelly23b, Hayes24}.

For our microlensing treatment, we assume a zero mean function $E[\delta\beta(t)] = 0$, treating the impact of microlensing as a perturbation around the BayeSN SED model. The choice of covariance function impacts the characteristic scale over which the function being modelled varies. In this work we use a Gibbs kernel \citep{Gibbs97}, first proposed for a treatment of microlensing in \citet{Hayes24}. The Gibbs kernel is non-stationary and allows the length scale parameter to vary with time. This quality is well suited for modelling microlensing as it allows for faster evolution as the SN photosphere crosses a stellar caustic and slower evolution elsewhere.

The Gibbs kernel describing the covariance of the Gaussian process between two phases $t$ and $t'$ is given by

\begin{equation}
\label{gibbs_equation}
    k^{\text{Gibbs}}_{\bm{\Lambda}}(t, t') = A^2\Bigg(\frac{2l(t)l(t')}{l^2(t)+l^2(t')}\Bigg)^{0.5}\exp\Bigg(-\frac{(t-t')^2}{l^2(t) + l^2(t')}\Bigg) 
\end{equation}
where $l(t)$ is a variable length-scale function given by
\begin{equation}
    l(t) = \lambda(1-p\phi_{(\tau_{\rm ML}, \eta)}(t)).
\end{equation}
The parameters $\bm{\Lambda} = \{A, \lambda, p, \tau_{\rm ML}, \eta\}$ are tuning parameters, which are not directly physically interpretable but would relate to the amplitude and size of a microlensing caustic as well as the SN ejecta velocity, since these would determine the size and timescale of the microlensing magnification. $\phi_{(\tau_{\rm ML}, \eta)}(t)$ is a Gaussian probability density function with mean $\tau_{\rm ML}$ and standard deviation $\eta$, while $A$ is an amplitude parameter. These GP parameters are independent for each image $i$, given that microlensing will impact each image differently. We refer to the set of microlensing parameters for each image collectively as $\bm{\Lambda}_{si} = \{ A_{si}, \lambda_{si}, p_{si}, \tau_{{\rm ML},si}, \eta_{si} \}$. The microlensing curve for each image $i$, $\delta\beta_{si}(t)$, is  modelled as a realisation of an independent Gaussian process:
\begin{equation}
    \delta\beta_{si}(t) \sim \mathcal{GP}(0, k^{\text{Gibbs}}_{\bm{\Lambda}_{si}}(t, t')).
\end{equation}
Given a set of observed phases of image $i$ of SN $s$, $\bm{t}_{si}$, the vector of microlensing curve values evaluated at these phases, $\bm{\delta\beta}_{si}$, thus has a multivariate Gaussian prior distribution:
\begin{equation}
    \label{eqn:gp_mvn}
    \bm{\delta\beta}_{si} \sim N(0, \bm{K}^{\text{Gibbs}}_{\bm{\Lambda}_{si}}(\bm{t}_{si},\bm{t}_{si}) ),
\end{equation}
where $\bm{K}^{\text{Gibbs}}_{\bm{\Lambda}_{si}}(\bm{t}_{si},\bm{t}_{si})$ is the covariance matrix with elements populated by the Gibbs kernel evaluated on all pairs of observed phases in $\bm{t}_{si}$.

It is important to note that for this work, we make the simplifying assumption of achromatic microlensing i.e. the same microlensing curve applies to all bands of a given image's light curve. As mentioned previously, for SNe Ia this approximation is valid for approximately the first 3 weeks after explosion, covering optical peak luminosity, but does not hold to later times \citep{FoxleyMarrable18, Goldstein18, Huber19, Huber21}. The SNTD method presented in \citet{Pierel19}, and applied for the SN H0pe analysis in \citet{Pierel24}, involves directly fitting colour curves which are insensitive to achromatic microlensing; the impact of chromatic microlensing is then considered as part of the systematic error budget. \citet{Grupa25} also analysed colour curves to remove the impact of achromatic microlensing. \citet{Hayes24} included an analytic, achromatic treatment of microlensing. A chromatic treatment of microlensing would be of interest to explore in future work but would add significant additional complexity to the model, though since the development of BayeSN-TD, \citet{Hayes25b} has developed a GP-based treatment of chromatic microlensing. In this work we test the robustness of time-delay estimation with an achromatic microlensing treatment to the impact of chromatic microlensing.

\subsection{Priors}
\label{priors}

In this section we detail the priors included when fitting multiply-imaged glSNe Ia with BayeSN-TD, which are outlined in Table \ref{priors_table}.

\begin{table}
    \centering
    \caption{Priors on BayeSN-TD parameters when fitting light curves of glSNe Ia. \newline $^* TN(\mu, \sigma^2, a, b)$ denotes a Truncated normal distribution with mean and variance $\mu$ and $\sigma^2$ prior to truncation, lower truncation bound $a$ and upper truncation bound $b$.}
    \begin{tabular}{cc}
    Parameter & Prior \\
    \hline
    $A_V$ & $A_V\sim \text{Exp}(0.32 \text{ mag})$\\
    $R_V$ & $R_V\sim TN^*(2.51, 0.65^2, 1.2, \infty)$\\
    $\theta_1$ & $\theta_1\sim N(0, 1)$ \\
    $A$ & $A\sim \text{Half}-N((0.1 \text{ mag})^2)$ \\
    $\lambda$ & $\lambda\sim U(10, 150)$ \\
    $p$ & $p\sim U(0, 1)$ \\
    $\tau_\text{ML}$ & $\tau_\text{ML}\sim U(-10, 85)$ \\
    $\eta$ & $\eta\sim U(1, 40)$ \\
    \end{tabular}
    \label{priors_table}
\end{table}

Each BayeSN model is defined over a specific phase range. For example, the models presented in \cite{M20}, \cite{T21} and \cite{W22} are all defined from $-10$ days to $+40$ days in the rest-frame relative to B-band maximum. However, given that time-of-maximum is not perfectly known a priori when fitting a SN light curve, it is important that the model has some ability to extrapolate beyond this phase range. This allows for the time-of-maximum to be sampled during light curve fitting without data falling in and out of phase coverage as the sampler explores possible $t_\text{max}$ values. In practice that corresponds to linear extrapolation of $W_0(t, \lambda_r)$, $W_1(t, \lambda_r)$ and $\epsilon^s(t,\lambda_r)$.

As a result, when fitting SN light curves with BayeSN, the following procedure is followed:

\begin{enumerate}
    \item Each SN requires a fiducial estimate for the time of maximum, $T_\text{max}$ - we will refer to this fiducial value as $\text{T}_\text{max}^\text{fid}$. This can be based on some simple algorithm, a previous SALT fit or a maximum a posteriori (MAP) estimate from the BayeSN model.
    \item This fiducial value is used to convert observer-frame MJDs into rest-frame phases relative to peak.
    \item Data is selected based on the rest-frame phase coverage of the model being used - data points outside of this phase range are discarded.
    \item When the light curve is fit, the parameter $t_\text{max}$ is treated as being a rest-frame shift to the fiducial value. We use a uniform prior on this shift such that $t_\text{max} \sim U(-10 \text{ days}, +10 \text{ days})$. This is equivalent to an observer-frame prior of $T_\text{max} \sim U(T_\text{max}^\text{fid} - 10\times(1+z_\text{hel}^s), T_\text{max}^\text{fid} + 10\times(1+z_\text{hel}^s))$, where $z_\text{hel}^s$ is the heliocentric redshift of a SN $s$.
    \item After fitting, the posterior distribution on $t_\text{max}$ can be used alongside the fiducial value $T_\text{max}^\text{fid}$ to obtain a posterior distribution on the observer-frame time-of-maximum.
\end{enumerate}

The prior window of 10 rest-frame days either side of $T_\text{max}^\text{fid}$ is imposed to prevent the model linearly extrapolating far beyond the range over which the model is defined. In particular, when allowed to extrapolate far beyond its specified range $\epsilon^s(t,\lambda_r)$ can exhibit some unphysical behaviour. This prior width far exceeds typical uncertainties on time-of-maximum. When using our model for time-delay estimation, we fit for a separate $T_\text{max}^i$ for each image $i$. 

\subsubsection{Priors on Microlensing}

The priors on the parameters in the Gibbs kernel which we use to model microlensing, described in Equation \ref{gibbs_equation}, are detailed in Table \ref{priors_table}. In general, we choose broad, uninformative priors for these kernel parameters. The exception to this is the case of the amplitude parameter $A$, for which we use a half-Normal prior with a scale factor of 0.1. This is chosen to reflect the typical scale of microlensing deviations while avoiding imposing a hard upper-limit to ensure that more extreme microlensing events can still be fit.

While these priors are ultimately arbitrary, when validating our model on simulations which incorporate a realistic treatment of microlensing we find that our model produces well-calibrated uncertainties and can capture deviations from a base SN Ia SED model caused by microlensing. This demonstrates the priors that we have used within our model are suitable.

\subsection{Full Posterior}

We now define the full posterior of the BayeSN-TD model. We define the complete set of parameters to be inferred as follows:
\begin{itemize}
    \item $\bm{\Theta}_s$: The set of parameters shared across all images, describing the physical properties of the SN and the impact of host galaxy dust extinction.
    \[
        \bm{\Theta}_s = \{ \theta_1^s, A_V^s, R_V^s, \epsilon^s \}
    \]
    \item $\bm{\Phi}_s$: The set of parameters that are specific to each of the $I$ lensed images. This is a collection of parameter sets, one for each image $i \in \{1, ..., I\}$.
    \[
        \bm{\Phi}_s = \{ \bm{\Phi}_{s1}, \bm{\Phi}_{s2}, ..., \bm{\Phi}_{sI} \}, \quad \text{where} \quad \bm{\Phi}_{si} = \{ T_{\text{max}}^{si}, \mu^{si}, \bm{\Lambda}_{si} \}
    \]
\end{itemize}

Here, $T_{\text{max}}^{si}$ is the observer-frame time of B-band maximum, $\mu^{si}$ is the apparent distance modulus (capturing both cosmological distance and magnification), and $\bm{\Lambda}_{si}$ represents the set of hyperparameters for the microlensing GP for image $i$.

Let $\bm{\mathcal{F}} = \{ \bm{\mathcal{F}}_1, ..., \bm{\mathcal{F}}_I \}$ be the full set of observed photometric light curve data for all images, $\bm{\mathcal{T}} = \{ \bm{\mathcal{T}}_1, ..., \bm{\mathcal{T}}_I \}$ be the full set of times of observation for all images, and $\bm{\mathcal{H}}$ be the set of fixed, pre-trained population hyperparameters from the base BayeSN model. Finally, $z_s$ is the spectroscopic redshift of SN $s$, which is used purely for time dilation and spectral redshifting, not for constraining distance.

The full joint posterior distribution for all unknown parameters conditional on the observed data is given below, factorised to explicitly show the contributions from each image and the shared properties of the source:
\begin{multline}
    P(\bm{\Theta}_s, \bm{\Phi}_s, \{\bm{\delta\beta}_{si} \}\mid \bm{\mathcal{F}}, \bm{\mathcal{T}}, \bm{\mathcal{H}}, z_s) \propto \\  \left[ \prod_{i=1}^{I} P(\bm{\mathcal{F}}_i \mid \bm{\mathcal{T}}_i, \bm{\Theta}_s, \bm{\Phi}_{si}, \bm{\delta\beta}_{si}, z_s) \times P(\bm{\delta\beta}_{si} \mid \bm{\Lambda}_{si}) \times P(\Phi_{si}) \right] \\ \times P(\bm{\Theta}_s \mid \bm{\mathcal{H}})
    \label{eq:bayeSN_TD_posterior}
\end{multline}
This expression comprises four key components: the data likelihood $P(\bm{\mathcal{F}}_i \mid \bm{\mathcal{T}}_i, \bm{\Theta}_s, \bm{\Phi}_{si},\bm{\delta\beta}_{si}, z_s)$, the GP prior (Eq. \ref{eqn:gp_mvn}) on the microlensing for each image $P(\bm{\delta\beta}_{si} \mid \bm{\Lambda}_{si})$, the prior on the parameters unique to each image $P(\bm{\Phi}_{si})$, and the prior on the shared SN parameters $P(\bm{\Theta}_s \mid \bm{\mathcal{H}})$. The prior terms are outlined in Section \ref{priors}. Assuming independent photometric measurements with Gaussian measurement uncertainties, the likelihood for the data of a single image $i$ of a SN $s$ is the product of the probabilities of each individual flux measurement. Note that we define the set of flux measurements $\mathcal{F}_i = \{ f_{ij} \}$. The likelihood for image $i$ is conditional on both the shared source parameters $\Theta_s$ and its own unique lensing parameters $\Phi_{si}$, such that:
\begin{multline}
    P(\bm{\mathcal{F}}_i \mid \bm{\mathcal{T}}_i, \bm{\Theta}_s, \bm{\Phi}_{si}, \bm{\delta\beta}_{si}, z_s) = \\ \prod_{j} \mathcal{N}\left(\hat{f}_{sij} \mid f_{sij}(T_{ij}, \delta\beta_{si}(t_{ij}), \bm{\Theta}_s, \bm{\Phi}_{si}, z_s), \sigma_{ij}^2\right)
\end{multline}
where $\hat{f}_{ij}$ and $\sigma_{ij}$ are the observed flux and its uncertainty for the $j$-th observation of image $i$, $T_{ij}$ is the time of this observation and $t_{ij}$ is the rest-frame phase of this observation. Including distance, magnification and microlensing effects, the model SED of each image becomes:

\begin{equation}
    F_{si}(t, \lambda_r) = S_s(t, \lambda_r) \times 10 ^{- 0.4[\mu_{si} + \delta\beta_{si}(t)]}
\end{equation}
where $S_s(t, \lambda_r)$ is the BayeSN model from equation \ref{bayesn_equation} and $\delta\beta_{si}(t)$ is the microlensing curve for image $i$ of supernova $s$, which is defined in magnitude space\footnote{Any overall magnification caused by microlensing will be degenerate with the combined distance-magnification parameter $\mu_{si}$, our microlensing treatments will capture relative changes in magnification during the light curve of each image.}. The model flux $f_{ij}$ is then defined by integrating this SED, redshifted to the observer-frame, through the filter of each observation $j$, with the likelihood evaluated in flux space. 
In this work, we evaluate the likelihood in flux space as we do not apply BayeSN-TD to consistently high signal-to-noise observations.

\subsection{Obtaining Posteriors on Time Delay}

One of the main goals when fitting multiply-imaged glSNe is to estimate the time delay between different images. BayeSN-TD does not directly fit for a time-delay parameter but does allow for posteriors on the time delay to be easily obtained. The model directly samples the time of maximum, implemented as a rest-frame shift from a fiducial value $\text{T}_\text{max}^\text{fid,i}$ as outlined in Section \ref{priors} - each image $i$ has an associated $t_\text{max}^i$ parameter. We can use the posteriors on $t_\text{max}^i$ to derive posteriors on $\text{T}_\text{max}^i$ by simply converting to observer frame,

\begin{equation}
    T_\text{max}^{i} = T_\text{max}^{\text{fid},i} + (1+z_\text{hel}^s) \times t_\text{max}^i
\end{equation}
for each posterior sample of $t_\text{max}^i$. After this, we obtain posteriors on the time delay between images $i$ and $k$, $\Delta T_{ik}$, by evaluating,

\begin{equation}
    \Delta T_{ik} = T_\text{max}^i - T_\text{max}^k
\end{equation}
for each step along our MCMC chains.

\subsection{Implementation of BayeSN-TD}

BayeSN-TD is a modification of the BayeSN code presented in \citet{Grayling24}, developed based on \textsc{numpyro} and \textsc{jax}. As a result, it shares the same advantages; the code is designed for GPU-acceleration and can perform Bayesian inference quickly and efficiently. Although the very limited samples of real observed glSNe mean that high computational performance is not essential---unlike regular SNe Ia---this does enable us to apply this complex model to large samples of simulated data to assess performance.

\section{Validation on Simulations}
\label{sec:sims}

We begin by assessing the performance of BayeSN-TD on simulated populations of lensed SNe Ia. To date, such simulations have generally used the SALT SED model for SNe Ia \citep{Guy07, Kenworthy21}. As a result, the application of our model to these simulations results in an inherent model misspecification. BayeSN has previously proven robust when inferring population-level properties from simulated data sets using SALT \citep{GraylingPopovic25}, but there will still be differences between the two. In our case, this is a valuable test given that in reality the empirical models we use to simulate and perform inference will never perfectly match the true properties of SNe Ia. By applying our model to these simulations, we can assess whether our results are robust when applied to data simulated from a different model.

\subsection{Roman Simulations}
\label{roman}

We first assess the performance of BayeSN-TD at recovering time delays from simulated glSNe from the \textit{Nancy G.\ Roman Space Telescope} presented in \cite{Pierel21}. These simulations were based on the "All-z" Roman SN observing strategy described in \cite{Hounsell18}, with a few modifications to reflect more recent survey updates. This pipeline used the extended SALT2 model presented in \citet{Pierel18} to simulate SNe Ia. These simulations incorporate the effect of microlensing based on 12 different microlensing maps; however, this treatment assumes achromatic microlensing similarly to the BayeSN-TD model. These simulations therefore provide an opportunity for testing our achromatic GP treatment of microlensing on realistic achromatic microlensing simulations.

\begin{figure}
\centering
\includegraphics[width = \linewidth]{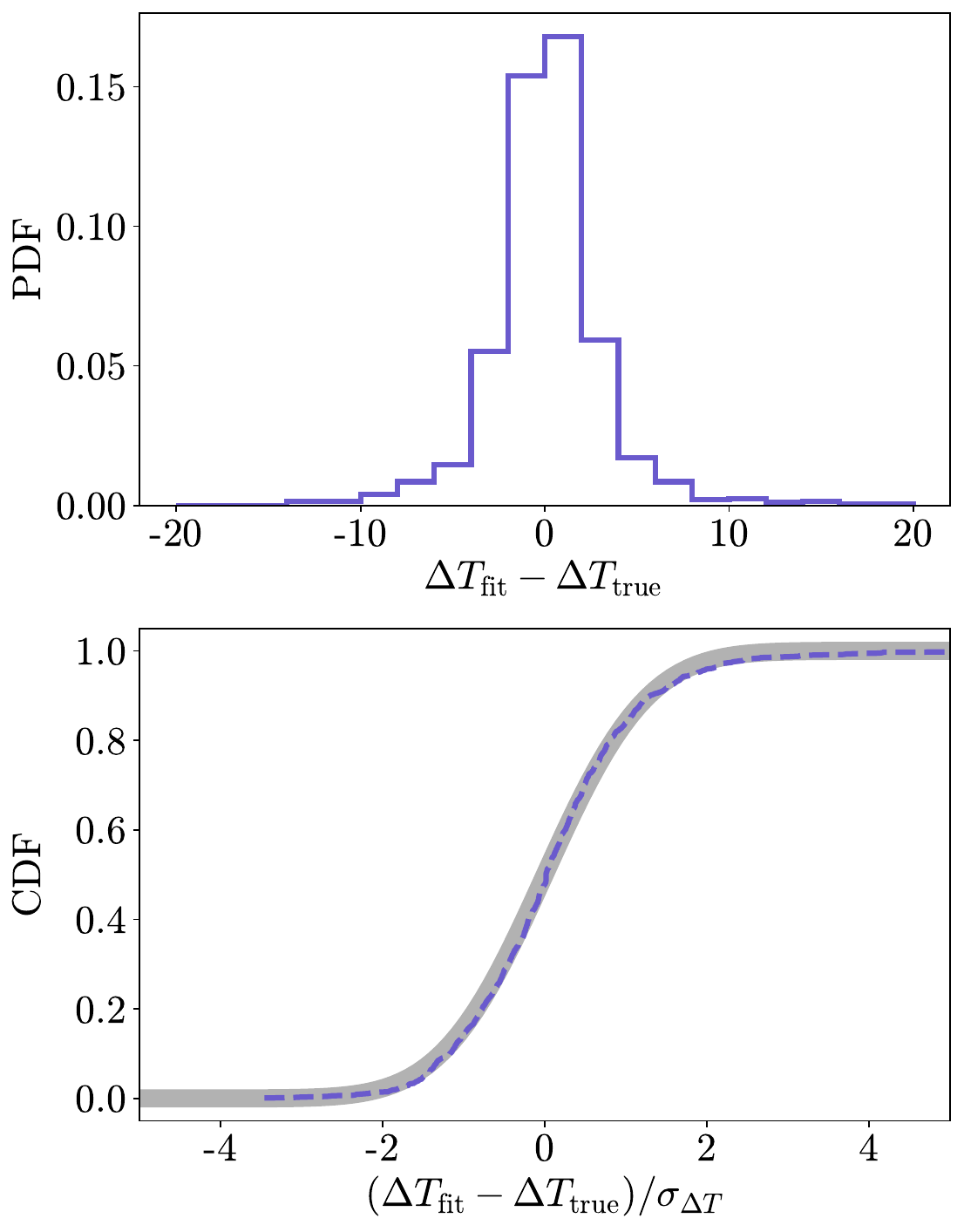}
\caption{\textbf{Upper:} Histogram showing distribution of time-delay residuals relative to true simulated values when applying BayeSN-TD to simulations of glSNe Ia observed by Roman presented in \citet{Pierel21}. \textbf{Lower:} Cumulative density of time-delay residual normalised by posterior uncertainties (dashed) shown alongside the expected cumulative density function for a Normal distribution represented by the shaded region. This demonstrates that BayeSN-TD produces well-calibrated uncertainties for these simulations.}
\label{roman_sim_plots}
\end{figure}

The results of time-delay recovery with BayeSN-TD are summarised in Table \ref{sim_results}. Despite the difference between the model used to simulate the light curves and the model used for inference, BayeSN-TD performs well at recovering the true simulated time delays. Overall, the bias between true and inferred time delays is 0.09 days, negligible when considering that across all simulated glSNe the mean of all posterior uncertainties is 3.07 days. The posterior uncertainties from BayeSN-TD are also well-calibrated - the true simulated time delays lie within the 68 and 95 per cent credible intervals in 67.7 and 93.5 per cent of cases respectively, quantities we will refer to as $f_{68}$ and $f_{95}$ hereafter. The top panel of Fig. \ref{roman_sim_plots} shows the distribution of residuals for inferred time delays, $\Delta T_\text{fit} - \Delta T_\text{true}$, and the bottom panel shows the cumulative density function of $(\Delta T_\text{fit} - \Delta T_\text{true}) / \sigma_{\Delta T}$ (the `pull') compared with a Gaussian CDF. This bottom panel further demonstrates that BayeSN-TD produces well-calibrated posteriors, with this distribution closely following a Gaussian.

\begin{figure*}
\centering
\includegraphics[width = \linewidth]{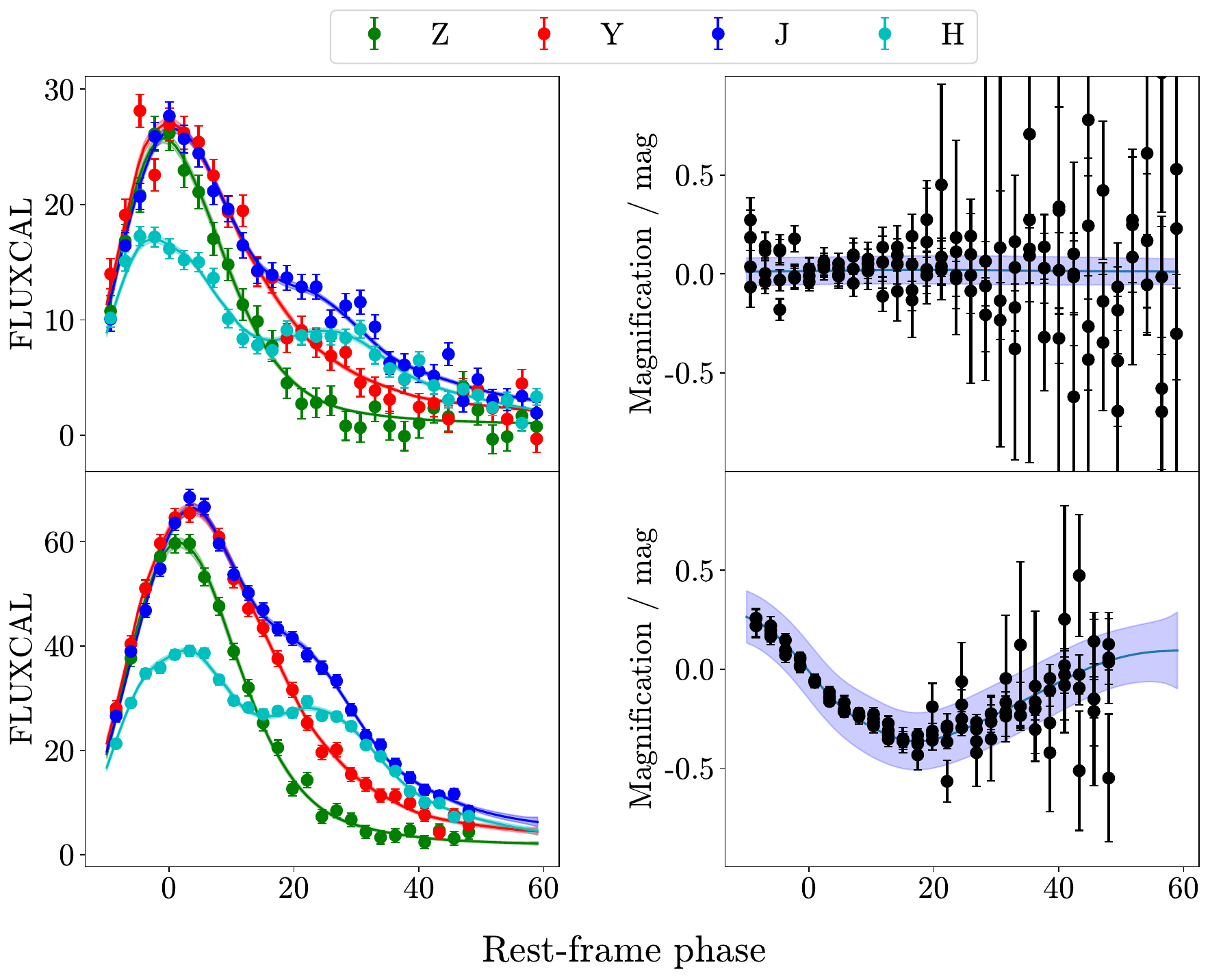}
\caption{\textbf{Left panels}: Simulated 2-image Roman glSN Ia light curve from \citet{Pierel24} along with associated BayeSN-TD fits. \textbf{Right panels}: Plotted data points represent simulated deviation from model light curves as a result of microlensing, with associated uncertainties from measurement noise as true simulated values post-microlensing, without noise, are not available. Plotted line and shaded region represent the posterior mean and standard deviation on microlensing from BayeSN-TD, demonstrating that with Roman simulations the model is able to constrain the deviation away from a typical SN Ia template as a result of microlensing. Note that these simulations, along with BayeSN-TD, assume achromatic microlensing.}
\label{Roman_example_good}
\end{figure*}

An example of a BayeSN-TD fit to one of the simulated Roman light curves is shown in Fig. \ref{Roman_example_good}. The left panels show the simulated photometry compared with the model fit for each image, while the right panels show the simulated achromatic microlensing curves compared with the posterior from BayeSN-TD. With this dataset, it was not possible to access the raw simulated microlensing curves - instead, the influence of microlensing can be determined by comparing the `observed' simulated magnitudes with the true simulated magnitudes before microlensing, albeit this is after the effect of measurement noise. This is why each point in the simulated microlensing curves has a corresponding uncertainty. Despite the mismatch between the models being used for simulation and for inference, it is clear that BayeSN-TD is able to closely match the simulated photometry. In addition, the lower right panel demonstrates that our model is able to successfully match the deviations from typical SEDs of SNe Ia as a result of microlensing. In some cases the posterior distribution will be centred around zero where the data does not provide any constraint on microlensing, such as in the upper right panel. However, the model is able to constrain cases of significant microlensing. Note that any overall magnification as a result of microlensing - a shift in the y-axis of the right panels - will be captured by BayeSN-TD's distance parameters, and these plots represent relative changes in microlensing magnification across the duration of the SN.

As mentioned above these simulations are based on the extended SALT2 model presented in \citet{Pierel18}, which extended the coverage of the default SALT2 template further into near-ultraviolet and near-infrared wavelengths using sophisticated extrapolation techniques. This extrapolation technique was applied for wavelengths less than 3500 \AA. It should be noted that these extrapolations were intended to enable simulations in these wavelength regimes, but not to make SALT2 capable of fitting light curves at these wavelengths. When applying BayeSN-TD to simulations based on rest-frame wavelengths significantly less than 3500 \AA, we found that differences between our BayeSN model and the extrapolated SALT2 model in this wavelength regime led to poor quality fits to simulated $Z$-band (F087 band) light curves. Differences between the models will be particularly prevalent at these wavelengths given SALT2 is based on extrapolation. To avoid this issue, we exclude $Z$-band data when fitting simulated glSNe where the observer-frame $Z$-band probes wavelengths bluer than 3000 \AA. This is done purely because BayeSN does not match SALT2 extrapolation in this region, and does not mean that BayeSN should not be applied to real data at these wavelengths.

\begin{table*}
    \centering
    \caption{Summary of performance of BayeSN-TD when applied to Roman simulations of glSNe Ia from \citet{Pierel24} which incorporate achromatic microlensing, along with performance when applied to LSST simulations from \citet{Arendse24} which incorporate chromatic microlensing. $f_{68}$ details the percentage of simulations where the true simulated value was within the 68 per cent credible interval of the posterior, while $f_{95}$ is the same but for the 95 per cent credible interval.}
    \begin{tabular}{cccccccc}
         Simulation & $N_\text{SN}$ & |$\Delta T_\mathrm{fit} - \Delta T_\mathrm{true}$| & |$\Delta T_\mathrm{fit} - \Delta T_\mathrm{true}$| & |$\Delta T_\mathrm{fit} - \Delta T_\mathrm{true}$| & Median & $f_{68}$ & $f_{95}$ \\
         & & < 1 day & < 3 days & < 5 days & $\Delta T_\mathrm{fit} - \Delta T_\mathrm{true}$ / days & \\
         \hline
         Roman & 1000 & 0.417 & 0.792 & 0.906 & 0.09 & 67.7\% & 93.5\% \\
         LSST & 1134 & 0.386 & 0.746 & 0.884 & -0.08 & 68.2\% & 90.2\%
    \end{tabular}
    \label{sim_results}
\end{table*}

\subsection{LSST Simulations}

\begin{figure}
\centering
\includegraphics[width = \linewidth]{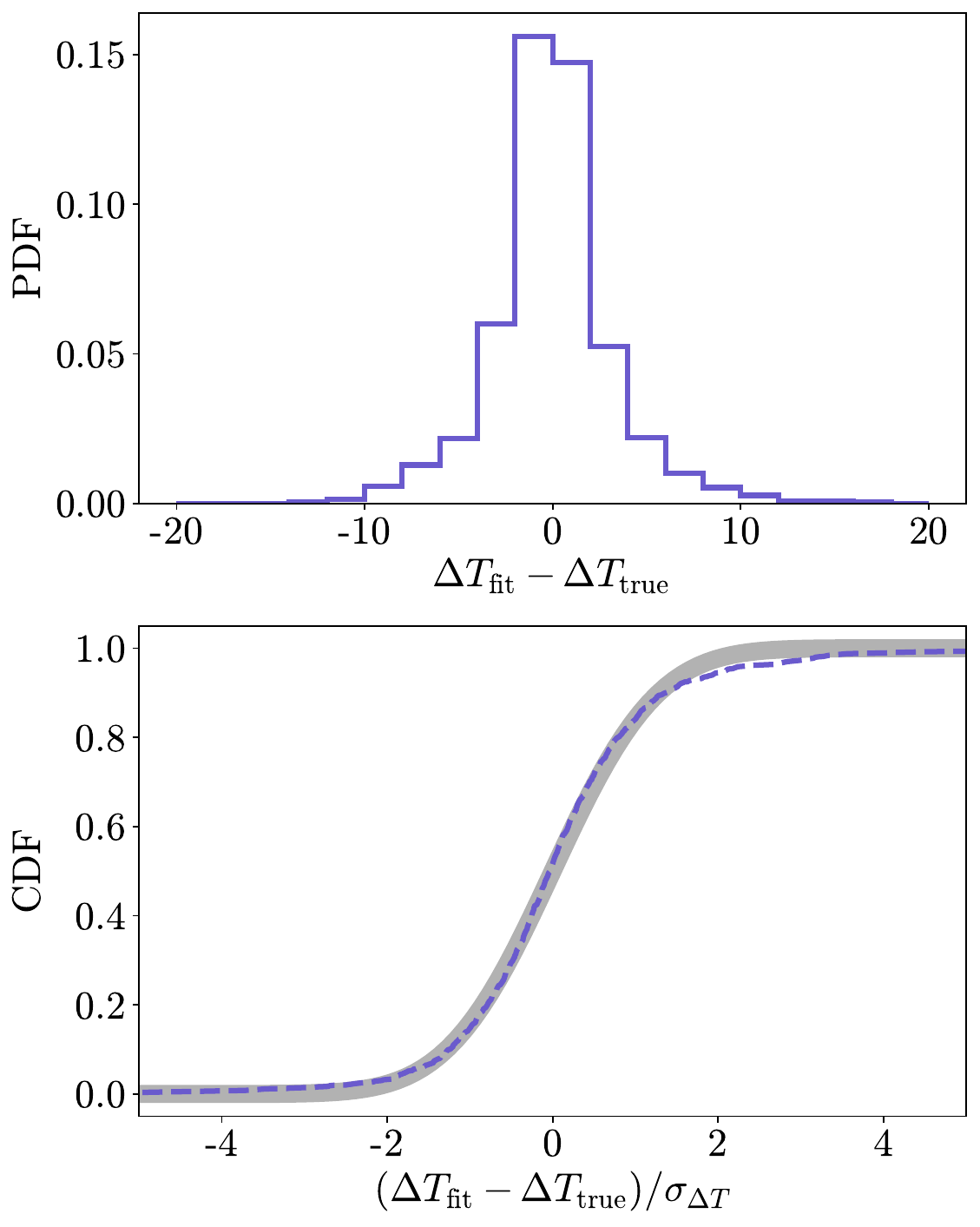}
\caption{As Fig. \ref{roman_sim_plots} but for LSST simulations of glSNe Ia presented by \citet{Arendse24} which incorporate a chromatic effect of microlensing.}
\label{lsst_sim_plots}
\end{figure}

We next explore realistic simulations of glSNe observed by LSST as presented in \cite{Arendse24}, using the `lensed Supernova Simulation Tool` (\textsc{lensedSST}). This pipeline uses the SALT3 model \citep{Kenworthy21} for the SEDs of glSNe Ia, simulated using \textsc{sncosmo} \citep{Barbary25}. 

Unlike the Roman simulations explored in Section \ref{roman}, these simulations do incorporate a chromatic treatment of microlensing. This provides an ideal opportunity for us to test whether our achromatic treatment of microlensing enables us to obtain robust time delay constraints of glSNe Ia which are impacted by chromatic microlensing. Microlensing is accounted for in the simulations using SN Ia explosion models from \textsc{ARTIS} \citep{Kromer09} combined with microlensing maps from \textsc{Gerlumph} \citep{Vernardos14a, Vernardos14b, Vernardos15}.

There are a number of simulated glSNe for which resolved\footnote{Cases where each image has individually-resolved photometry as opposed to blended photometry of both images combined.} photometry was not available for both images. These simulations aimed to provide a general, realistic data set for analysis of lensed supernovae including those without resolved photometry. However, in our case BayeSN-TD is aimed for application to SNe with resolved photometry of each image such as SN H0pe. For this analysis, we select only simulated SNe with at least 10 data points for each image, across all bands. This data quality cut requires only a small number of data points per photometric band. Out of 5000 total simulated glSNe Ia available, we apply BayeSN-TD to 1134 objects.

The results of time-delay recovery with BayeSN-TD for these LSST simulations are summarised in Table \ref{sim_results}. Compared with the Roman simulations with achromatic microlensing, the inclusion of chromatic microlensing in the simulations makes only a small impact to model performance. Most notably, this does not lead to a bias in the inferred time delays - the median deviation from the truth across all 1134 simulated SNe that were fit was just -0.08 days, negligible compared to the 3.08 day mean posterior uncertainty across all SNe. Even with chromatic microlensing, the uncertainties remain well-calibrated with $f_{68}=68.2\%$. There is a small decrease in $f_{95}$ compared with the Roman simulations, from 93.5\% to 90.2\%, suggesting that chromatic microlensing is causing a larger fraction of outliers. This is unsurprising, and it remains reassuring that there is a negligible overall bias and $f_{95}$ remains close to 95\%. Fig. \ref{lsst_sim_plots} is similar to Fig. \ref{roman_sim_plots} but shows results for these LSST simulations rather than the Roman simulations. The bottom panel of Fig. \ref{lsst_sim_plots} further demonstrates that BayeSN-TD can provide well-calibrated uncertainties on time delays; the distribution of $(\Delta T_\text{fit} - \Delta T_\text{true}) / \sigma_{\Delta T}$ closely follows a Gaussian with just a small number of outliers in the tails.

\begin{figure*}
\centering
\includegraphics[width = \linewidth]{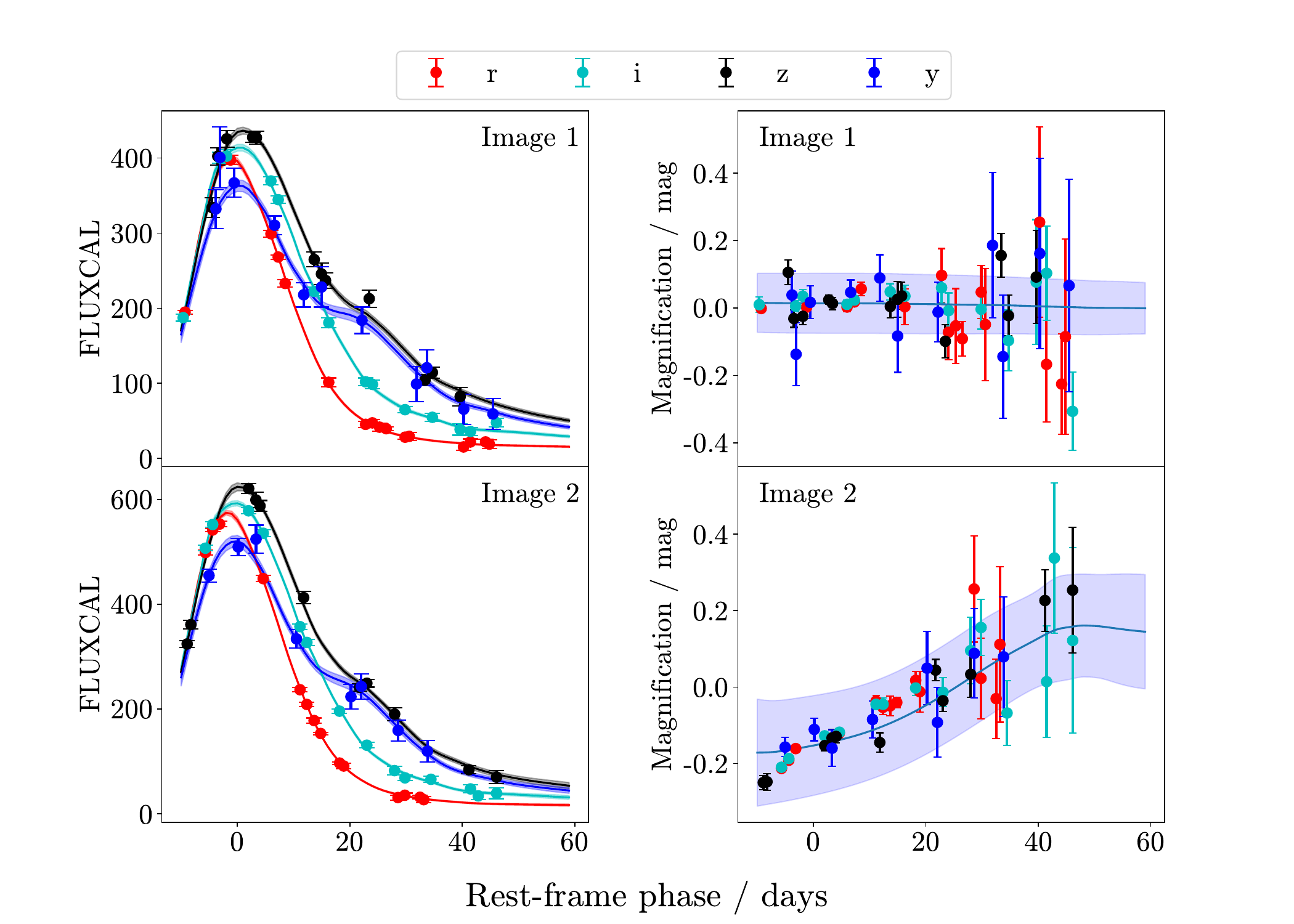}
\caption{\textbf{Left panels}: Simulated 2-image LSST glSN Ia light curve from \citet{Arendse24} along with associated BayeSN-TD fits. \textbf{Right panels}: Plotted data points represent simulated deviation from model light curves as a result of microlensing, with associated uncertainties from measurement noise as true simulated values post-microlensing, without noise, are not available. These simulations include chromatic microlensing, therefore different filters are differently impacted. Plotted line and shaded region represent the posterior mean and standard deviation on microlensing from the achromatic treatment included in BayeSN-TD.}
\label{LSST_example}
\end{figure*}

An example of a BayeSN-TD fit to one of these simulated LSST light curves is shown in Fig. \ref{LSST_example}. The left panels show the model fits along with the simulated photometry for each image, while the right panels compare the posterior distributions on microlensing curves with the impact of microlensing on the simulated data. As with the Roman simulations, the plotted data shows the deviation from the model on the simulated light curves as a result of microlensing, incorporating the effect of measurement noise. These plots show the different impacts of microlensing in each band along with the posterior distribution we obtain on the microlensing curve from our achromatic treatment.

\begin{figure*}
\centering
\includegraphics[width = \linewidth]{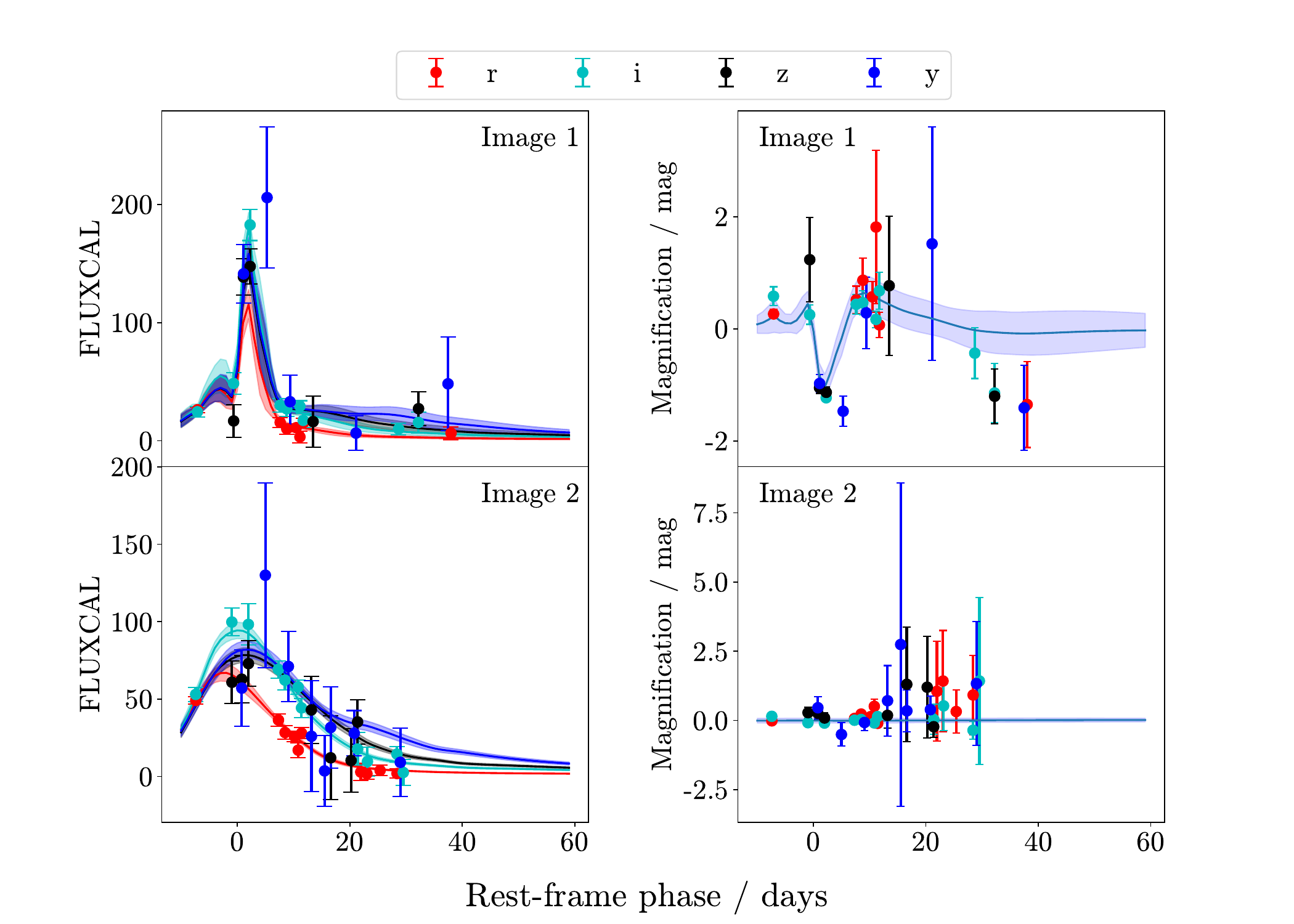}
\caption{As Fig. \ref{LSST_example} but for a particularly extreme example of microlensing. In this case, BayeSN-TD is able to identify that the peak of this light curve is driven by microlensing rather than the luminosity of the SN, and recover the true time delay within 1 day.}
\label{LSST_example_extreme}
\end{figure*}

This example demonstrates that our GP treatment of microlensing is able to capture deviations around the template of typical SNe Ia, even considering the mismatch between the SALT models used for the simulations and the BayeSN model used for inference. The lower right panel shows the posterior distribution on microlensing curve closely tracing the impact of microlensing on the simulated light curve. A further example is shown in Fig. \ref{LSST_example_extreme}, where the BayeSN-TD model has been able to capture an extreme microlensing event and infer that the very brightest points of the SN light curve are driven by microlensing rather than SN luminosity. This perfectly demonstrates the ability of our GP-based microlensing treatment to realistically capture the varied impact of microlensing. Overall, our achromatic treatment of microlensing seems to roughly average over the impact of microlensing in each band.

\section{Application to SN H0pe}
\label{sec:h0pe}

Having established that our model is able to robustly infer time delays for SNe Ia which are impacted by microlensing in Section \ref{sec:sims}, we now apply BayeSN-TD to real photometry of SN H0pe to estimate time delays and magnifications. 

\subsection{Light Curve Fits}

We begin by fitting the observed photometry of SN H0pe with BayeSN-TD, presented in Table 2 of \citet{Pierel24}. As discussed in Section \ref{priors}, BayeSN-TD fits for time-of-maximum of each image relative to some fiducial peak phase. The prior on $t_\text{max}$ is then a uniform distribution 10 rest-frame days either side of this fiducial peak phase for each image. Considering the high redshift of SN H0pe, this means that in the observer-frame the priors on the peak MJD of each image are broad, uninformative uniform distributions such that

\begin{equation*}
\begin{aligned}
    \text{MJD}_\text{max,2a} \sim 59924 + U(-27.8, 27.8)\hspace{2pt} \\
    \text{MJD}_\text{max,2b} \sim 60033 + U(-27.8, 27.8)\hspace{2pt} \\
    \text{MJD}_\text{max,2c} \sim 59989 + U(-27.8, 27.8).
\end{aligned}
\end{equation*}

\begin{figure}
\centering
\includegraphics[width = \linewidth]{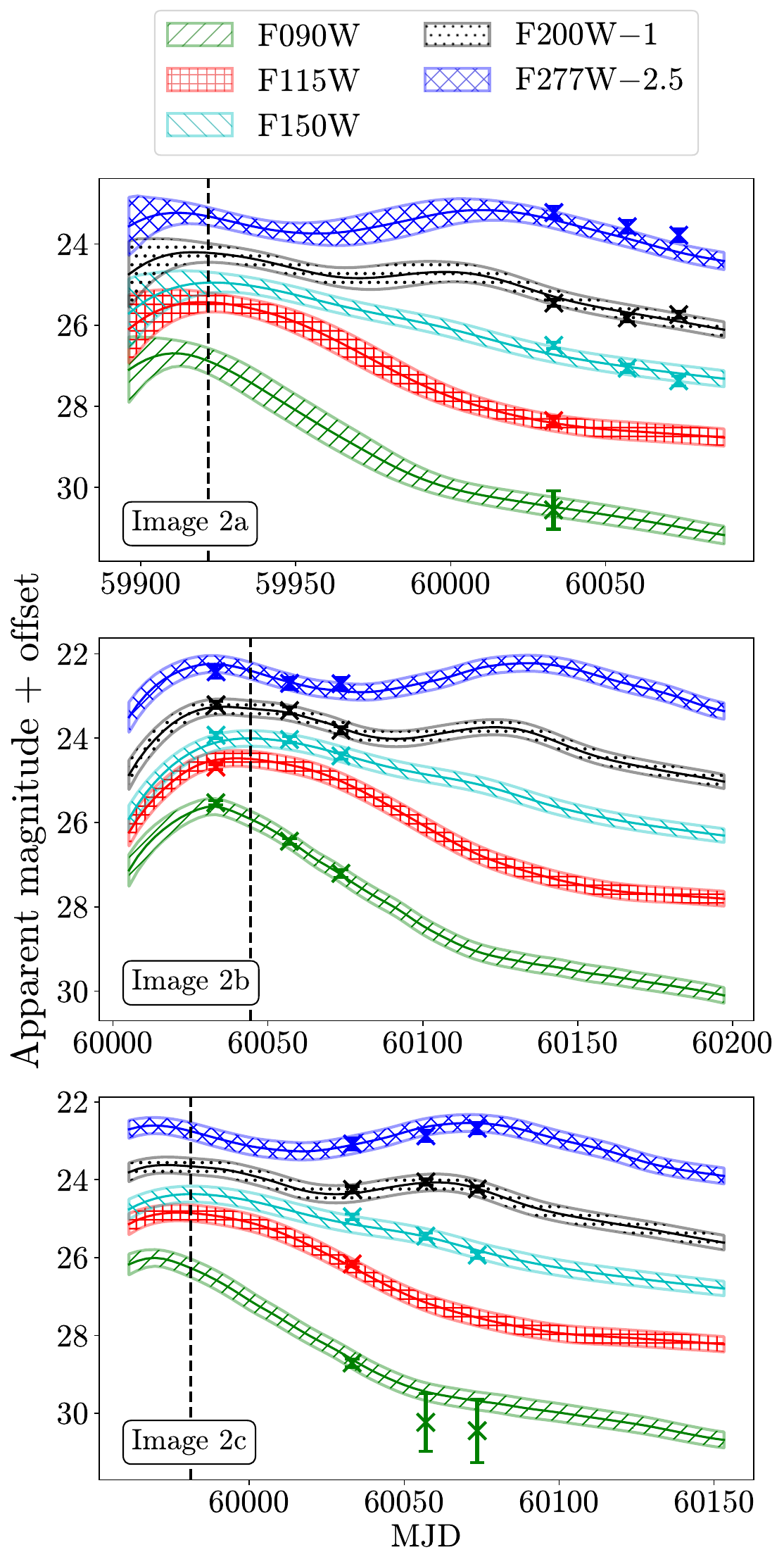}
\caption{BayeSN-TD fits to each image of SN H0pe. Lines and hashed regions represent the posterior mean and standard deviation on the light curve fits. Dashed vertical lines represent the phase of rest-frame B-band peak luminosity.}
\label{h0pe_fit_plot}
\end{figure}

\begin{figure*}
\centering
\includegraphics[width = \linewidth]{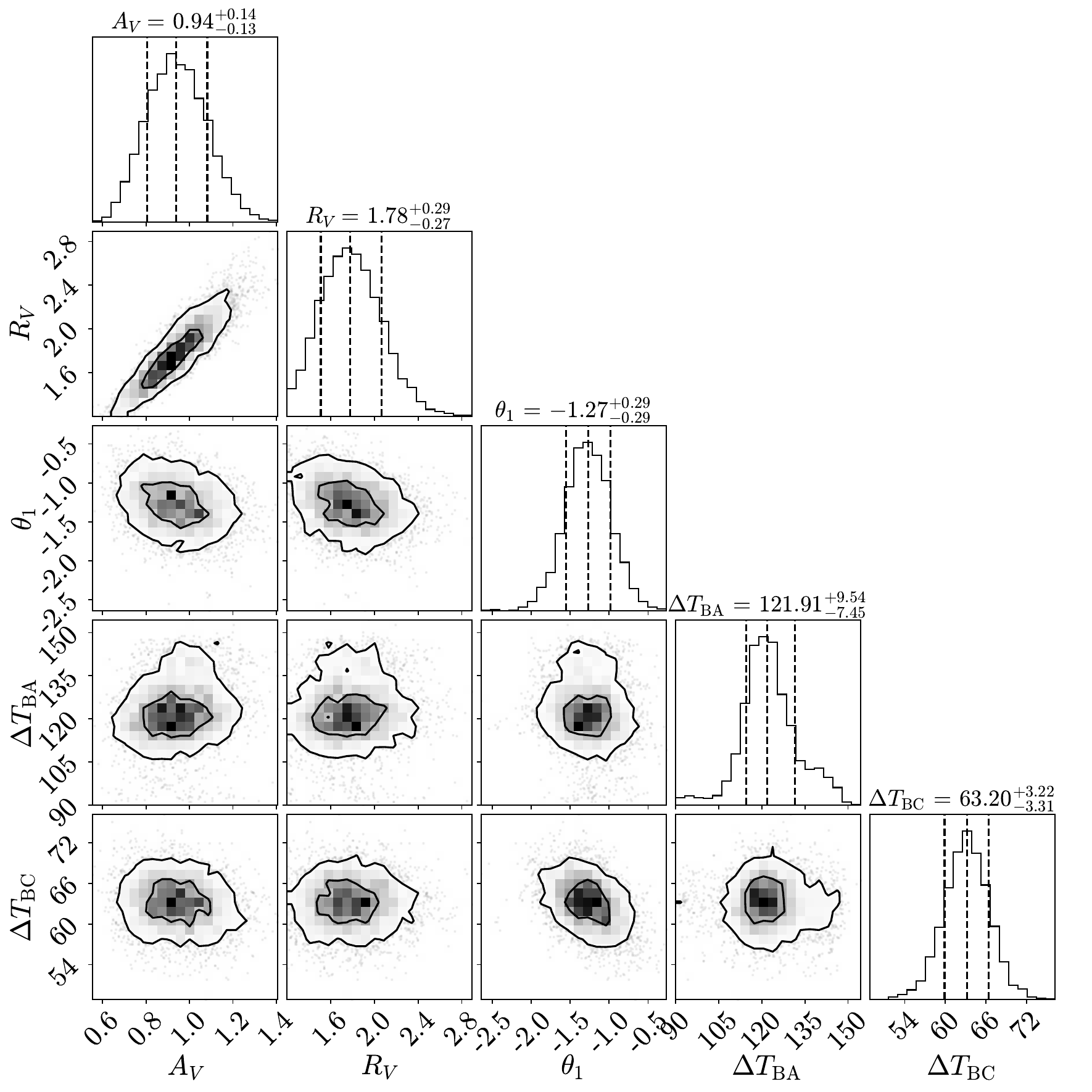}
\caption{Corner plot showing joint and marginal distributions on $A_V$, $R_V$, $\theta_1$ and time delays $\Delta T_{BA}$ and $\Delta T_{BC}$ when fitting SN H0pe with BayeSN-TD. Quoted values represent posterior medians and 68 per cent credible intervals. For visual clarity we have excluded $\beta_A$, $\beta_B$ and $\beta_C$ from this plot, though the full joint posterior was used for $H_0$ inference.}
\label{h0pe_corner}
\end{figure*}

Our BayeSN-TD fits to the light curve of SN H0pe are shown in Fig. \ref{h0pe_fit_plot}, and fit parameters are shown in Table \ref{h0pe_fit_params}. Fig. \ref{h0pe_corner} shows joint posteriors on BayeSN model parameters and time delays. For clarity we exclude magnifications from this figure; we see some covariance between $\Delta T_\text{BA}$ and $\beta_A$, with weaker covariances between other time delays and magnifications. We infer $\theta=-1.27\pm0.29$, which corresponds to a B-band 15-day decline $\Delta m_{15,\text{B}}\approx0.91$ mag. We find that SN H0pe has a large amount of host-galaxy dust reddening, with $A_V=0.95\pm0.14$ and a relatively low $R_V=1.80\pm0.28$. A low value of $R_V$ is fairly typical for more highly-reddened SNe Ia \citep[e.g.][]{TM22, Burns14, Amanullah14}. Unsurprisingly, as shown in Fig. \ref{h0pe_corner}, there is a large degree of covariance between $R_V$ and $A_V$.

\begin{table}
    \centering
    \caption{Summary of parameter estimates inferred for SN H0pe when fit with BayeSN-TD, including time delays $\Delta T_{ij}$ and flux space magnifications $\beta_i$. Values quoted as $X\pm Y$ represent posterior means and standard deviations, while values quoted as $X^{+Y}_{-Z}$ represent posterior medians and 68 per cent credible intervals.}
    \begin{tabular}{cc}
    Parameter & Value  \\
    \hline
    $\theta$ & $-1.27\pm0.29$ \\
    $A_V$ & $0.95\pm0.14$ \\
    $R_V$ & $1.80\pm0.28$ \\
    $\Delta T_{BA}$ & $121.9^{+9.5}_{-7.5}$ days \\
    $\Delta T_{BC}$ & $63.2^{+3.2}_{-3.3}$ days \\
    $\beta_A$ & $2.38^{+0.72}_{-0.54}$ \\ 
    $\beta_B$ & $5.27^{+1.25}_{-1.02}$ \\ 
    $\beta_C$ & $3.93^{+1.00}_{-0.75}$ \\ 
    \end{tabular}
    \label{h0pe_fit_params}
\end{table}

\subsubsection{Microlensing of SN H0pe}

\begin{figure}
\centering
\includegraphics[width = \linewidth]{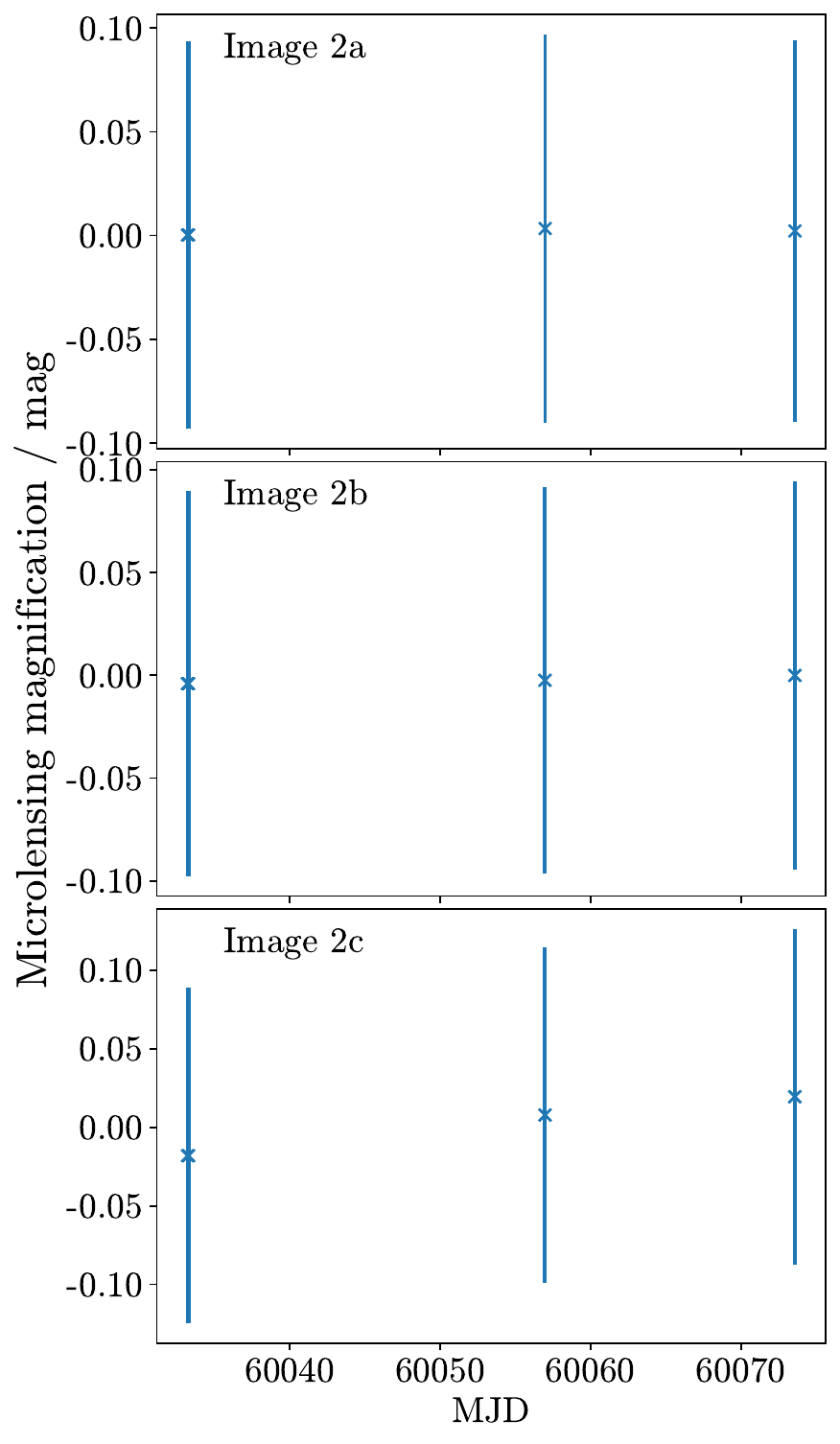}
\caption{Posterior mean and standard deviation of microlensing magnification, shown for each image against MJD.}
\label{h0pe_ml_constraints}
\end{figure}

Given that BayeSN-TD incorporates a GP-based treatment of microlensing, we can also examine our posteriors on microlensing magnification to see if the model predicts significant microlensing for the light curve of SN H0pe. Fig. \ref{h0pe_ml_constraints} shows the posterior mean and standard deviation for the microlensing magnification at each epoch of photometry for each image. For images A and B, this line is effectively flat, consistent with no time-varying impact of microlensing. Note that this does not rule out that the observed light curves are influenced by microlensing, simply showing that we cannot constrain its impact with the available data. Image C seems to qualitatively show a weak upward trend in magnification between the available epochs of photometry, but considering the size of this change relative to the posterior uncertainties we cannot make a strong conclusion. Unlike with the very well-sampled Roman light curves shown in Section \ref{roman}, for SN H0pe we cannot obtain good constraints on microlensing from the available data by considering deviations away from typical SN Ia templates.

\subsubsection{Differences with Previous Results}
\label{td_differences}

There are some notable differences between some of the parameters inferred in this work and those from \citet{Pierel24}, most notably the time delay between images B and C, $\Delta T_{BC}$. There are a number of methodological differences between these two works, which are:

\begin{enumerate}
    \item In this work we apply the model directly to photometry, which contains colour information. In contrast, \citet{Pierel24} applies SNTD to fit the observations specifically in colour space rather than in light curve space.
    \item We marginalise over an achromatic treatment of microlensing when fitting SN light curves. \citet{Pierel24} does not explicitly consider the impact of microlensing on top of the SN Ia SED model; instead, by fitting colours rather than photometry, the method employed by \citet{Pierel24} is insensitive to achromatic microlensing. 
    \item We use a different BayeSN model, training a new model with extended phase coverage and more SNe in the training set; \citet{Pierel24} used a phase-extended version of the BayeSN model presented in \citet{W22}.
    \item We incorporate $\epsilon^s(t,\lambda_r)$, detailed in Section \ref{bayesn_equation}, within the model when fitting, marginalising over the distribution of residual intrinsic SN colours. \citet{Pierel24} does not marginalise over this distribution when fitting, instead considering the impact of $\epsilon^s(t,\lambda_r)$ as part of the systematic error budget.
\end{enumerate}

To investigate what is driving the difference in $\Delta T_{BC}$, we did explore modifying our analysis to remove some of these differences; we repeated our analysis using the same BayeSN model as in \citet{Pierel24}, disabling our GP microlensing treatment and fixing $\epsilon^s(t,\lambda_r)=0$ rather than marginalising over the distribution of residual intrinsic scatter. However, we found that these modifications did not make significant differences to our results and did not reconcile the differences between $\Delta T_{BC}$ inferred this work and \citet{Pierel24}. As discussed in Section \ref{h0pe_sims}, we apply BayeSN-TD to simulations of SN H0pe which include chromatic microlensing and find that it is able to provide robust time delays and magnifications. As a result, when inferring $H_0$ in Section \ref{H0_constraints} we directly use our inferred time-delay and magnification constraints and do not consider an additional uncertainty contribution resulting from these differences. Overall, the exact cause of this discrepancy remains uncertain; the full reanalysis of SN H0pe to be presented in \textcolor{blue}{Agrawal et al. in prep.}, with higher accuracy photometry from new template images, will illuminate the root cause.

\subsection{Magnification}

As mentioned previously, one of the key advantages of using strongly-lensed SNe Ia for $H_0$ inference is our ability to standardise them and probe absolute magnifications, which can break the mass sheet degeneracy and provide further information about the lens. Please note, we hereafter refer to the absolute magnification of an image $i$ as $\beta_i$. BayeSN fits for a distance modulus $\mu^s$ jointly with all other parameters in a light curve fit. For BayeSN-TD, in practice the magnification will be degenerate with distance - in this case we fit for a distance modulus $\mu^s_i$ for each image $i$, where this single parameter captures both effects. One approach to estimate the magnification would be to evaluate the distance modulus at the redshift of the SN under an assumed cosmological model, $\mu_\text{cosmo}(z_s)$, and obtain a posterior on the magnification of each image by evaluating absolute magnification $\beta_i =10^{-0.4(\mu^s_i - \mu_\text{cosmo}(z_s))}$, giving a flux space magnification for each step along the chain. However, this relies on an assumed cosmological model including an assumed value of $H_0$. \citet{Pierel24} instead estimates a cosmology-independent magnification by comparing the apparent magnitude of SN H0pe to predictions of the apparent magnitude of non-lensed SNe Ia at the redshift of SN H0pe based on fits to the Pantheon+ sample \citep{Brout22}. We follow a similar approach here to infer an absolute magnification without relying on an assumed cosmology. 

To do this, we fit all 39 SNe Ia with a redshift $z>1$ in the Pantheon+ \citep{Brout22} sample with the BayeSN model we present in this work to estimate $\mu^s$ for each SN $s$. For this redshift range ($1<z<2.3$), we expect a relation that is approximately linear between $\mu(z)$ and $\log_{10}(z)$. We therefore fit for a linear relationship, $\mu(z) = \mu_{z=1.783} + b[\log_{10}(z) - \log_{10}(1.783)]$, based on the redshifts and fitted photometric distances for all SNe Ia in this redshift range, to make a prediction for what the expected distance modulus at a redshift of 1.783 should be. When fitting this, we find $\hat{\mu}_{z=1.783}=45.51\pm0.12$ for the assumed BayeSN $H_0$ value of 73.24 \kmsmpc. This uncertainty is a statistical uncertainty based on Bayesian linear regression implemented using \textsc{numpyro}. \citet{Pierel24} considers a variety of alternative methods to predict the apparent magnitude of a SN Ia at $z=1.783$, including a second-order polynomial, a GP and a kinematic expansion model \citep{Riess22}, taking a systematic uncertainty based on the standard deviation across these three methods of 0.14 mag. We consider our choice of a linear model in log space for this redshift range ($1<z<2.3$) to be reasonable and solely focus on the straight-line method, considering that our uncertainty is comparable to that from the method used in \citet{Pierel24}.

With this established, we estimate the absolute magnification of each image by computing the posterior distribution of $\beta_i = 10^{-0.4(\mu^\text{H0pe}_i - \mu_{z=1.783})}$. We do this numerically by evaluating
\begin{equation}
\label{mag_eqn}
    \beta_{ni} =10^{-0.4(\mu^\text{H0pe}_{ni} - N_n(45.51,0.12^2))}
\end{equation}
for each step $n$ along our MCMC chains, where $i$ denotes each image.  Let $\mu^\text{H0pe}_{ni}$ denote the posterior sample of the photometric distance modulus of the $i$th image of SN H0pe at step $n$ of the chain, and $N_n(45.51,0.12^2)$ denotes a Gaussian random variate with mean $45.51$ mag and standard deviation $0.12$ mag, with an independent random sample drawn for every $n$th step along the chain. The resulting samples $\beta_{ni}$ enable us to estimate the posterior distribution of the absolute magnification of each image which incorporates the uncertainty in $\mu_i^\text{H0pe}$ as well as the uncertainty in the estimated expected distance modulus at the redshift of SN H0pe, $\mu_{z=1.783}$, whilst marginalising over the intrinsic scatter of SN H0pe as well as in the sample of unlensed SNe Ia in this redshift range.

Note that as all of the photometric distance moduli in Eq. \ref{mag_eqn} were inferred using the same model, the reference value used within BayeSN cancels out, hence our magnification values are independent of $H_0$.

As discussed in Section \ref{h0pe_sims}, there are a few additional sources of systematic uncertainty impacting our estimates of magnification. We modify these posteriors to incorporate this additional uncertainty in our results. Our final marginal posteriors on the magnification of each image are $\beta_A = 2.38^{+0.72}_{-0.54}$, $\beta_B=5.27^{+1.25}_{-1.02}$ and $\beta_C=3.93^{+1.00}_{-0.75}$, also reported in Table \ref{h0pe_fit_params}; these values represent posterior medians and 68 per cent credible intervals. Please note that the joint posterior distribution of all magnifications and time delays was used to inference $H_0$. 

\subsection{Simulations of SN H0pe}
\label{h0pe_sims}

We have demonstrated in Section \ref{sec:sims} that BayeSN-TD performs well when applied to simulations of glSNe, including those simulated with chromatic microlensing. However, photometric data available for SN H0pe is very sparse compared with simulations of LSST and Roman. As a result, it remains important to validate the performance of our model on simulations more representative of SN H0pe.

We create a new set of simulations of SN H0pe using the method and framework outlined in Section 5 of \citet{Pierel24}. The only difference in our case is that we use the new, 85-day BayeSN model presented in this work as the basis for the simulations. These simulations include a realistic, chromatic treatment of microlensing. As discussed previously, microlensing is effectively achromatic for SNe Ia in the first 3 weeks after explosion but chromatic effects become increasingly prominent at later times. While image 2b of SN H0pe is likely in this `achromatic phase', images 2a and 2c cover later rest-frame phases and may be impacted by chromatic microlensing. As our BayeSN-TD model incorporates only an achromatic treatment of microlensing, chromatic effects may lead to biased constraints on time delays and magnifications from SN H0pe. We therefore apply BayeSN-TD to these simulations to assess whether our assumption of achromatic microlensing leads to biased constraints or posteriors which lack Frequentist coverage.

We produce 1000 simulations of SN H0pe-like glSNe including the impact of chromatic microlensing, residual intrinsic chromatic scatter and photometric uncertainties, analogous to the `Combined' simulations presented in Section 5.3 of \citet{Pierel24} and fit them using BayeSN-TD in order to assess whether our model produces well-calibrated posterior distributions when these effects are present. We create these simulations with a range of $A_V$, $R_V$ and $\theta$ values such that $\theta\sim N(0, 1)$, $A_V\sim U(0, 1.5)$ (to cover a wide range of reddened values, given the highly-reddened nature of SN H0pe) and $R_V\sim N(2.51, 0.65^2)$ following the constraints of \citet{Grayling24} (though as discussed in \citet{Pierel24} time-delay inference is not significantly impacted by $R_V$). The simulated light curves have the same cadence and noise as the real observations of SN H0pe. We simulate a range of time-delays close to the values inferred in this work and in \citet{Pierel24}, from $\sim110-150$ days for $\Delta T_{BA}$ and $\sim47-63$ days for $\Delta T_{BC}$\footnote{\textbf{Our simulation code follows the approach of that used in \citet{Pierel24} including the simulated phases, which is why this range for $\Delta T_{BC}$ is mostly below that inferred in this work, though the value that we infer is still covered.}}. We also simulate these events over a wide range of magnifications which cover our inferred values, in mag space drawn from a Gaussian distribution $N(0, 2^2)$ for each image independently. Chromatic microlensing effects are applied on top of the simulated BayeSN light curves; the details of the microlensing simulations used for this are outlined in Section 5.2.2 of \citet{Pierel24}. In brief, these simulations convolve magnification maps for each SN image with SN Ia light profiles from four theoretical models \citep{Suyu20, Huber21}, following the procedure of \citet{Huber19}. For each simulated glSN we apply a random realisation of one of these microlensing curves.

One difference between this work and the analysis presented in \citet{Pierel24} is that \citet{Pierel24} adapts the BayeSN model by incorporating it within the SNTD framework to estimate time delays based on colour curves, while BayeSN-TD is a modification of the \textsc{numpyro} BayeSN code presented in \citet{Grayling24}. The main practical difference relates to treatment of BayeSN's residual intrinsic chromatic scatter term $\epsilon^s(t,\lambda_r)$, as described in Section \ref{bayesn_equation}. BayeSN-TD incorporates this term within the model and marginalises over the population distribution of residual intrinsic scatter, whereas SNTD excludes $\epsilon^s(t,\lambda_r)$ within the fitting process and instead considers its potential impact on time delays/magnifications as a systematic uncertainty. Given that we are marginalising over the distribution of $\epsilon^s(t,\lambda_r)$, unlike \citet{Pierel24} we do not consider its impact as a possible systematic.

The recovery of true simulated time-delays and magnifications for these 1000 simulations are shown in Table \ref{h0pe_sim_coverage}. While BayeSN-TD produced well-calibrated posteriors for simulated LSST and Roman glSNe in Section \ref{sec:sims}, it is noticeable that our posteriors for SN H0pe-like simulations are in fact overly conservative; the 68 and 95 per cent credible regions cover the truth for more than 68 and 95 per cent of our simulations for all parameters except $\Delta T_{BC}$. In the case of $\Delta T_{BC}$, the 68 and 95 per cent credible regions cover the truth for 65.3 and 92.4 per cent of simulations respectively. These posteriors are very slightly overconfident but close to the expected Frequentist coverage. The main difference between these simulations of SN H0pe compared to the LSST and Roman simulations in Section \ref{sec:sims} is that SN H0pe has far fewer observations and less phase coverage; this lack of data explains the difference in posterior coverage between the different simulations. We consider it reassuring that our BayeSN-TD model produces underconfident, rather than overconfident, constraints when analysing glSNe with sparse data coverage.

Overall, these results show that our analysis is not significantly impacted by chromatic microlensing - including this effect in our simulations does not lead to significantly overconfident posteriors. With this in mind, we opt not to add an additional systematic contribution to our uncertainties in time delays and magnification as a result of microlensing and instead leave our posterior distributions on time delays unaltered.

There are, however, a few additional sources of systematic uncertainty which impact constraints on magnification that are not included in these simulations. One of these is the impact of millilensing, additional lensing caused by dark matter subhalos associated with the cluster and halos along the line of sight. Millilensing was found to contribute an additional $\sim10\%$ uncertainty in inferred magnifications for SN Refsdal \citep{Kelly23b}. We use the expected impacts on SN H0pe magnifications from millilensing presented in \citet{Pierel24}, as detailed in Section 5.4 of that work using techniques from \citet{Gilman19, Gilman20}. We incorporate this as an additional source of uncertainty to our posteriors.

We do see some small biases in recovery of inferred parameters, particularly of the time delay between images B and C. \citet{Kelly23b} and \citet{Pierel24} correct their inferred posteriors for these biases by applying small shifts to inferred values on real data. Such bias corrections are reliant on having very realistic simulations and would in reality need to have their own associated uncertainties. As we have demonstrated that the biases are small compared to our posterior uncertainties, and that the posteriors have good coverage of the truth for these simulations, we opt not to apply bias corrections in our analysis. Were we to apply them, they would lead to minor shifts in our parameter estimates which are small compared to the uncertainties. 

\begin{table}
    \centering
    \caption{Recovery of simulated time-delays and magnifications across 1000 simulated SN H0pe-like glSNe simulated with BayeSN, showing the median offset and the fraction of cases where the truth lies in the 68 per cent and 95 per cent credible intervals of our posteriors, $f_{68}$ and $f_{95}$.}
    \begin{tabular}{cccc}
    Parameter & Median offset & $f_{68}$ & $f_{95}$ \\
    \hline
    $\Delta T_{BA}$ & -0.22 days & 80.4\% & 98.3\% \\
    $\Delta T_{BC}$ & 0.59 days & 65.3\% & 92.4\% \\
    $\beta_A$ & -0.035 & 73.3\% & 97.3\% \\
    $\beta_B$ & -0.022 & 74.5\% & 97.7\% \\
    $\beta_C$ & 0.023 & 72.8\% & 97.6\% \\
    \end{tabular}
    \label{h0pe_sim_coverage}
\end{table}

\section{Inferring $H_0$}
\label{H0}

Having inferred time delays and magnifications for SN H0pe, we next apply these to estimate $H_0$ by combining these constraints with the lens modelling presented in \citet{Pascale25}. Please note that we do not develop upon any of these lens models in this work, simply aiming to combine the constraints from BayeSN-TD with existing models. \citet{Agrawal25} explored the consistency of these lens models with magnification constraints from SN H0pe. In the near future, new templates of the lensing system of SN H0pe will enable more precise photometry, which will lead to improved time-delay constraints to be presented in \textcolor{blue}{Agrawal et al. in prep.} along with associated $H_0$ constraints.

\subsection{Time-delay Cosmography}

We briefly outline here how the time delay can be used to constrain $H_0$. Full details are presented in the first $H_0$ analysis of SN H0pe presented in \citet{Pascale25}, building on the framework of \citet{Kelly23a} and using the time-delay constraints from \citet{Pierel24}.

The time delay between an individual source at position $\beta$ and a corresponding lensed image at angular position $\theta$ can be expressed as:

\begin{equation}
t(\boldsymbol{\theta}, \boldsymbol{\beta}) = \frac{1 + z_l}{c} \frac{D_l D_s}{D_{ls}} \left[ \frac{1}{2} (\boldsymbol{\theta} - \boldsymbol{\beta})^2 - \psi(\boldsymbol{\theta}) \right]
\end{equation}

where $\psi(\boldsymbol{\theta})$ is the lensing potential at the observed image position $\boldsymbol{\theta}$, $z_l$ is the redshift of the lensing cluster and $D_l$, $D_s$ and $D_{ls}$ respectively are the angular diameter distances to the lens, source and between the lens and source. Alternatively, considering the time delay between any two sets of images $i, j$ of the same system, the time delay between them is:

\begin{equation}
\Delta T_{i,j}(\boldsymbol{\theta}_i, \boldsymbol{\theta}_j) = D_{\Delta T} \, \Delta \tau_{i,j}(\boldsymbol{\theta}_i, \boldsymbol{\theta}_j, \boldsymbol{\beta})
\end{equation}

where $\tau(\boldsymbol{\theta}, \boldsymbol{\beta}) = \frac{1}{2} (\boldsymbol{\theta} - \boldsymbol{\beta})^2 - \psi(\boldsymbol{\theta})
$ is the Fermat potential \citep{Schneider85, Blandford86} and $D_{\Delta T} = \frac{1 + z_l}{c} \frac{D_l D_s}{D_{ls}}$ is the time delay distance, which depends on the angular diameter distances and is therefore dependent on cosmology \citep{Refsdal64, Schneider92, Suyu10}. Of particular relevance to this work, $D_{\Delta T}$ is inversely proportional to $H_0$ and is very weakly dependent on other cosmology parameters \citep{Bonvin17}. As a result, combining a measured set of time delays with a model of the lens potential enables constraints on $H_0$.

For the analysis presented in \citet{Pascale25}, seven different teams constructed models for the cluster lensing system to predict the lensing potentials at the positions of the images of SN H0pe. Each team constructed a lens model blinded from each other and without knowledge of any of the measured time delay and magnification constraints. Each lens model leads to a separate constraint on $H_0$, which are then combined to yield the final presented $H_0$ value. As detailed in Section 7 of \citet{Pascale25}, when combining these to obtain the final posterior distribution on $H_0$ these individual constraints are weighted by how each lens model is able to produce the measured time delays (and, optionally, measured magnifications). Effectively, this penalises lens models that are not able to reproduce the properties of the system measured using SN H0pe and lends additional weight to those more successful in producing the observations.

As detailed above, in this work we re-use these same lens models, produced before the first analysis of SN H0pe was unblinded, with no further modification.

\subsection{Constraints on $H_0$}
\label{H0_constraints}

We now present the constraints on $H_0$ we obtain when combining the lens models and weighting scheme presented in \citet{Pascale25} with the time delay and magnification constraints obtained in this work with BayeSN-TD. As discussed in Section \ref{sec:tdonlyH0} and \citet{Agrawal25}, there are significant differences in the magnifications predicted by the blinded lens models used for this analysis and inferred from SN H0pe. The focus of this work is time-delay inference specifically; development of updated lens models or alternative treatments for lens model weighting falls beyond the scope. \textcolor{blue}{Pascale et al. 2025, in prep} will present updated lens models which will address the magnification discrepancies.

\subsubsection{Constraints from Time Delays and Magnifications}

\begin{figure}
\centering
\includegraphics[width = \linewidth]{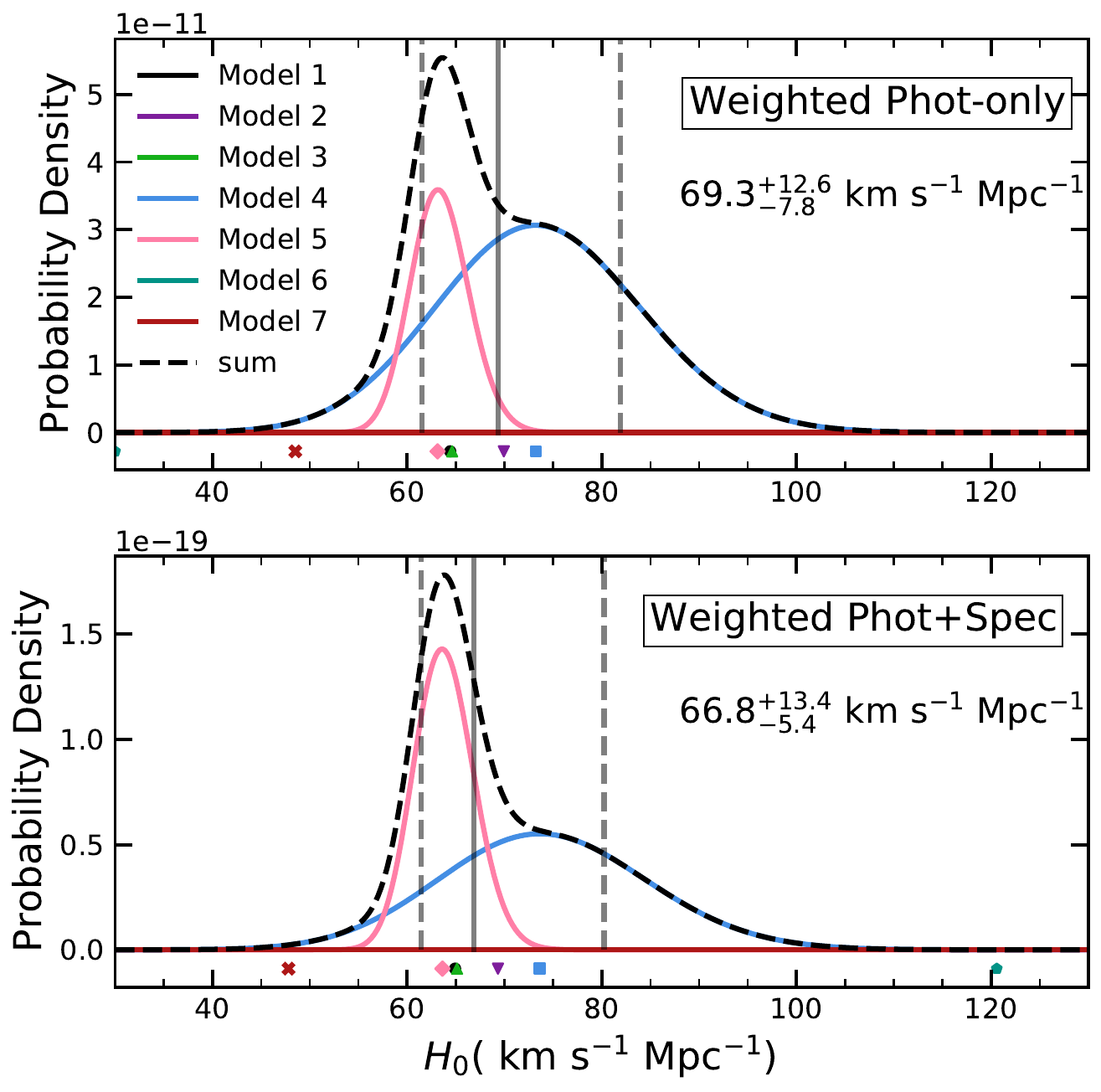}
\caption{\textbf{Top}: Posterior on $H_0$ when combining our constraints on time-delays and magnifications with each lens model from \citet{Pascale25}, along with the overall posterior on $H_0$ when combining those from each lens model weighted by how well they produce the measured time-delays and magnifications of SN H0pe. \textbf{Bottom}: The same as the top panel but combined with constraints on time-delays and magnifications from spectroscopic analysis of SN H0pe, presented in \citet{Chen24}.}
\label{weightedH0}
\end{figure}

We first consider constraints on $H_0$ which incorporate our inferred time delays and magnifications, the `weighted phot-only' case. In this case, the seven different lens models are weighted according to how close their predicted magnifications and time delays are to our measurements; please see Appendix \ref{lens_weights} for further details. The top panel Fig. \ref{weightedH0} shows the posterior on $H_0$ for this case, also showing the constraints on $H_0$ for each lens model and how they are each weighted to contribute to the final posterior. For this case, we obtain $H_0=69.3^{+12.6}_{-7.8}$ \kmsmpc. This value is lower but still consistent with that obtained in \citet{Pascale25}, courtesy of our shift in time-delays relative to \citet{Pierel24}. This value is consistent with both results obtained from early-Universe measurements from the Cosmic Microwave Background \citep{Planck20} and distance ladder-based constraints from local-Universe measurements \citep[e.g.][]{Reiss22, Li25, Freedman24}, and does not allow for comment with regard to the `Hubble tension'. For lensed SNe to make a more definitive conclusion in this debate, further analysis is required - please see Section \ref{H0_discussion} for more discussion of this.

The lower panel of Fig. \ref{weightedH0} shows the `weighted phot+spec' case. This is the posterior we obtain on $H_0$ when combining our constraints from photometric data in this work with those from analysing spectroscopic data of SN H0pe to infer time delays as presented in \citet{Chen24}. In this case, we obtain $H_0=66.8^{+13.4}_{-5.4}$ \kmsmpc, obtaining a slight increase in precision from using both photometric and spectroscopic data.

\subsubsection{Time-delay Only Constraints}
\label{sec:tdonlyH0}

We next consider constraints on $H_0$ derived purely from the time delays, not factoring in constraints on magnification. In this case, the seven lens models are instead weighted when inferring $H_0$ according to how well they predict the ratio of the two time delays of SN H0pe (see Section 7.1 of \citet{Pascale25} for further details).

\begin{figure}
\centering
\includegraphics[width = \linewidth]{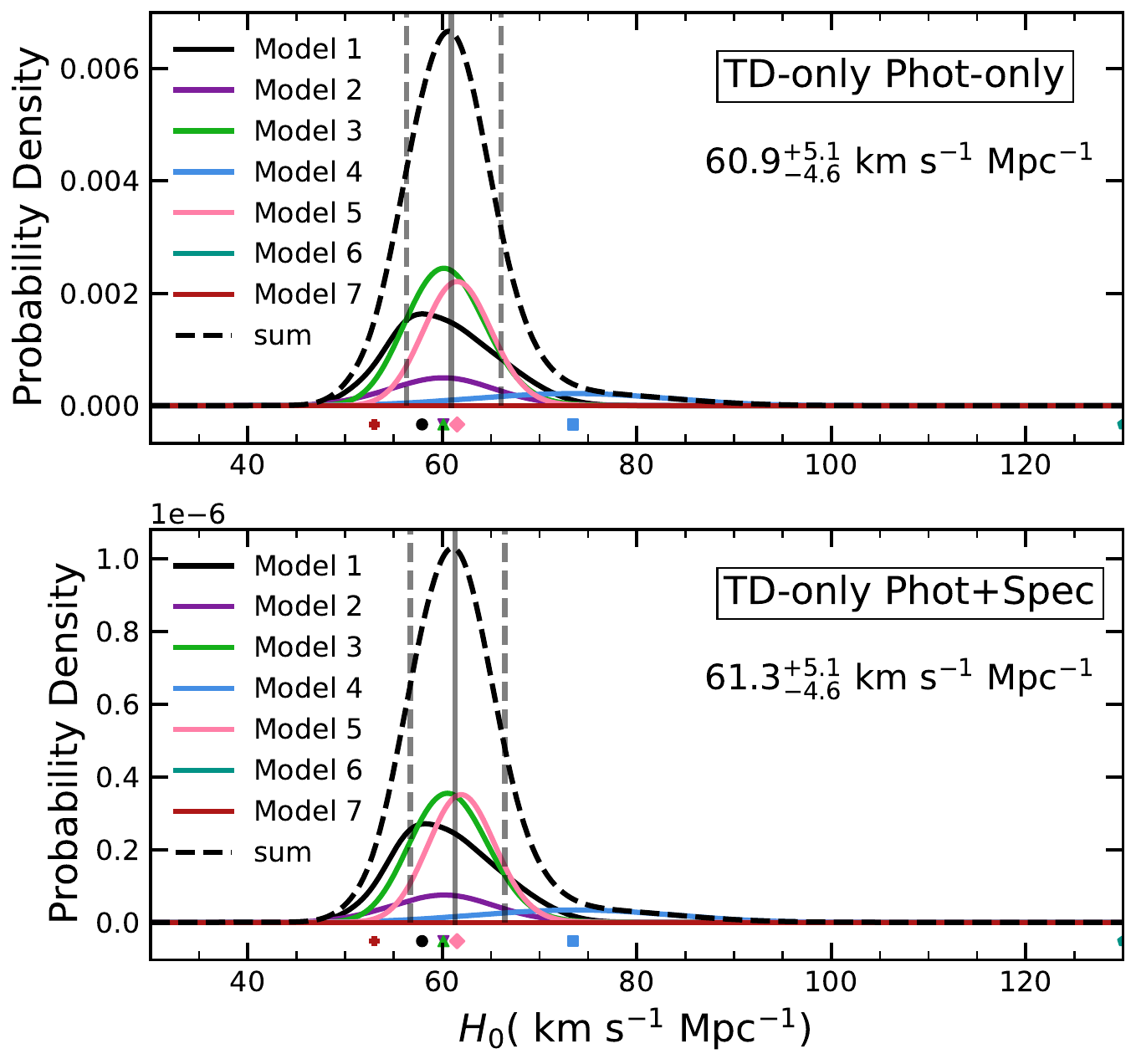}
\caption{\textbf{Top}: Posterior on $H_0$ when combining our constraints on time-delays alone with each lens model from \citet{Pascale25}, along with the overall posterior on $H_0$ when combining those from each lens model weighted only by how well they produce the measured time-delays of SN H0pe, not incorporating magnifications. We stress that these results should be treated with caution, given that they do not take into account how well each lens model agrees with our inferred magnification of SN H0pe. \textbf{Bottom}: The same as the top panel but combined with constraints on time-delays from spectroscopic analysis of SN H0pe, presented in \citet{Chen24}.}
\label{tdonlyH0}
\end{figure}

In this `TD-only' case, we infer $H_0=60.9^{+5.1}_{-4.6}$ \kmsmpc, though we emphasise that we would caution against drawing any conclusions relating to the `Hubble tension' from this result. The unique advantage of using glSNe Ia is that their standardisable nature allows for constraints on absolute magnifications; as discussed in \citet{Agrawal25}, lens models of SN H0pe seem to be consistently overestimating absolute magnifications of SN H0pe. In the TD-only case, each lens model is only weighted by how well they are able to reproduce the ratio of observed time delays and without any consideration of the model magnifications \citep{Oguri03}. This method seems to lead to a concordance across lens models relative to the `weighted phot-only' case in that many predict similar values of $H_0$, leading to a more precise constraint. However, for each lens model the predicted time delays are correlated with the predicted magnifications, meaning that incorporating magnification information shifts the inferred $H_0$ for each model. Lens models which predict the time-delay ratios well do not necessarily predict the magnifications well, hence looking at time-delay ratios alone does not capture a full physical picture of the system. \citet{Pascale25} comments on how absolute magnifications provide additional leverage to break the mass sheet degeneracy between lens models and weight them in $H_0$ inference. Considering that as a whole the lens models overestimate the absolute magnifications of SN H0pe \citep{Agrawal25}, we consider the `weighted phot-only' and `weighted phot+spec' results which weight the lens models by the magnification constraints to be our primary results.

\subsubsection{Discussion and Future Prospects}
\label{H0_discussion}

In this work, we have applied BayeSN-TD to the photometry of SN H0pe presented in \citet{Pierel24} to estimate time delays and magnifications of SN H0pe, and combined these with the lens modelling presented in \citet{Pascale25} to obtain constraints on $H_0$. Our primary inferred value is $H_0=69.3^{+12.6}_{-7.8}$ \kmsmpc, or $H_0=66.8^{+13.4}_{-5.4}$ \kmsmpc when combined with spectroscopic information, from weighting the series of lens models of SN H0pe presented in \citet{Pascale25} by how well they reproduce our inferred time delays and magnifications of this system. For discussion of constraints obtained by considering the time delays alone, please see Section \ref{sec:tdonlyH0}.

Throughout this work, we have demonstrated the ability of BayeSN-TD to obtain robust constraints on time delays for glSNe Ia while marginalising over, or even inferring, deviations from a standard SN Ia SED template as a result of microlensing. By applying our model to SN H0pe, we have also demonstrated the practical utility of this code by obtaining constraints on the time delays and magnifications of this system in order to infer $H_0$.

Unfortunately, with the available data for SN H0pe we are unable to obtain precise enough constraints on $H_0$ to make a meaningful conclusion regarding the Hubble tension. However, future prospects for inference of $H_0$ are very promising. As previously mentioned, since the publication of SN H0pe photometry in \citet{Pierel24} new templates of the system have been taken which will allow for more precise SN photometry and improved lens modelling. Updated analysis of SN H0pe using the updated photometric data will be presented by \textcolor{blue}{Agrawal et al. in prep.}. Multiply-imaged, gravitationally lensed SNe will be a vital cosmological probe over the next decade, given our expectations for an order-of-magnitude increase in the sample of known glSNe from  LSST \citep[e.g.][]{Goldstein19, Wojtak19, SainzDeMurieta23, SainzDeMurieta24, Arendse24, Bronikowski25}, and the expected constraints on $H_0$ this will enable \citep{Huber19, Suyu20}. BayeSN-TD can play a pivotal role in the analysis of these objects.

\section{Conclusions and Future Work}
\label{sec:conclusions}

We present BayeSN-TD, a modified version of the probabilistic SN Ia SED model BayeSN designed for application to multiply-imaged, gravitationally lensed SNe Ia. BayeSN's implementation as a hierarchical Bayesian model makes it naturally suited for such an application, where some parameters of the model are common across the images of a glSN while some latent parameters are independent for each image. We develop upon the BayeSN model by also incorporating a GP-based treatment of microlensing, which allows for time-dependent deviations from the SED template. In this way, BayeSN-TD can infer time-delays and magnifications which marginalise over the potential impact of microlensing. We implement this using a Gibbs kernel, assuming an achromatic impact of microlensing. In this work, we detail the BayeSN-TD model and validate it through application to simulations of glSNe Ia. Our findings can be summarised as follows:

\begin{itemize}
    \item Motivated by the modelling requirements of SN H0pe and similar analyses, we train a new BayeSN model with coverage out to 85 rest-frame days post-peak, far beyond the +40 day coverage of the models presented in \citet{T21, M20, W22} or the extended 50 day version of the \citet{W22} model which was applied in \citet{Pierel24}. This model is trained on $U$-band data to push the wavelength coverage of the model as blue as possible, and extends as far in phase as the Hsiao template (which BayeSN uses as a base template) allows. We make this model publicly available as part of the public BayeSN code.
    \item We apply BayeSN-TD to Roman simulations of glSNe Ia from \citet{Pierel21}, which incorporate the impact of achromatic microlensing. Despite the simulations being based on SALT, an entirely separate SED model for SNe Ia, we find that BayeSN-TD is able to infer robust time delays with well-calibrated posteriors with good coverage of the true simulated values. This demonstrates that our GP-based treatment of microlensing allows for realistic marginalisation over the potential impact of microlensing on inferred time-delays. In addition, with the well-sampled light curves expected from Roman, BayeSN-TD is able to constrain the time-varying impact of microlensing. Overall, our results from applying BayeSN-TD to these simulations strongly validate its performance, albeit these results do not test how chromatic microlensing may impact performance.
    \item We next apply BayeSN-TD to LSST simulations of glSNe from \citet{Arendse24}, which instead incorporate the impact of chromatic microlensing. In this case, we have two differences between the simulations and model used for inference; the simulations are based on SALT and feature chromatic microlensing, while BayeSN-TD uses BayeSN and assumes achromatic microlensing. Nevertheless, BayeSN-TD again performs well and obtains robust time-delay constraints and well-calibrated uncertainties with only a small impact on performance compared to the Roman simulations.
    \item While our assumption of achromatic microlensing has proven reasonable for the application of BayeSN-TD through testing on simulations, it would be preferable to remove this simplifying assumption and instead incorporate a chromatic treatment of microlensing. Considering the additional complexity involved and the good performance of BayeSN-TD, we leave this to future work, but this would help to improve the model.
    \item We also apply BayeSN-TD to simulations of SN H0pe. While these simulations are based on BayeSN, they also include the impact of chromatic microlensing. We find that for this case BayeSN-TD is able to produce well-calibrated or even under-confident constraints on time-delays and magnifications.
    \item Having established the ability of BayeSN-TD to obtain robust constraints on properties of lensed glSNe Ia, we then apply it to photometric data of SN H0pe presented in \citet{Pierel24}. We obtain constraints on the time delays between images B and A, and B and C, of $\Delta T_{BA}=121.9^{+9.5}_{-7.5}$ days and $\Delta T_{BC}=63.2^{+3.2}_{-3.3}$ days. For the absolute magnifications $\beta$ of each image, we infer $\beta_A = 2.38^{+0.72}_{-0.54}$, $\beta_B=5.27^{+1.25}_{-1.02}$ and $\beta_C=3.93^{+1.00}_{-0.75}$ - note that these are linear (flux) space magnification factors. Our inferred $\Delta T_{BA}$ is consistent with the value obtained in \citet{Pierel24}, however our inferred $\Delta T_{BC}$ is $\sim15$ days larger than the value obtained in \citet{Pierel24}. As outlined in Section \ref{td_differences}, there are a number of methodological differences between these two analyses, however many of them do not seem to impact our results and we are unable to identify the specific cause of this difference. Nevertheless, the final value for $H_0$ remains in statistical agreement with that inferred by the time delays from \citet{Pierel24}, making this difference only critical to understand for the future study with updated photometry (\textcolor{blue}{Agrawal et al. in prep.}), where the time-delay, and corresponding $H_0$, uncertainty is expected to improve substantially.
    \item We combine our constraints on time-delays and magnifications of SN H0pe with the lens modelling presented in \citet{Pascale25} to obtain constraints on $H_0$. Using our time delays and magnification to weight the seven different lens models, we infer $H_0=69.3^{+12.6}_{-7.8}$~\kmsmpc and slightly more precise constraints of $H_0=66.8^{+13.4}_{-5.4}$~\kmsmpc when combining this with constraints from the spectroscopic analysis presented in \citet{Chen24}. While these results are not yet precise enough to draw a meaningful conclusion with regard to the Hubble tension, newly available templates of the host and lensing system of SN H0pe will enable more precise photometry and improved lens modelling that will enable more precise constraints on $H_0$. \textcolor{blue}{Agrawal et al. in prep.} will present analysis of SN H0pe using this new data.
    \item BayeSN-TD obtains robust constraints of glSNe Ia properties when applied to simulations, and has been used to obtain constraints on $H_0$ from SN H0pe. We expect that glSNe Ia will become a major cosmological probe going forward, and BayeSN-TD can play a key role in their analyses.
\end{itemize}

\section*{Acknowledgements}

The authors thank Brenda Frye and Suhail Dhawan for helpful discussion. MG and KSM are supported by the European Union’s Horizon 2020 research and innovation programme under ERC Grant Agreement No. 101002652 (BayeSN; PI K.\ Mandel) and Marie Skłodowska-Curie Grant Agreement No. 873089 (ASTROSTAT-II). This research was supported in part by the Munich Institute for Astro, Particle and BioPhysics (MIAPbP) which is funded by the Deutsche Forschungsgemeinschaft (DFG, German Research Foundation) under Germany's Excellence Strategy – EXC-2094 – 390783311. ST was supported by funding from the European Research Council (ERC) under the European Union's Horizon 2020 research and innovation programmes (grant agreement no. 101018897 CosmicExplorer) and by the research project grant `Understanding the Dynamic Universe' funded by the Knut and Alice Wallenberg Foundation under Dnr KAW 2018.0067. JDRP is supported by NASA through Einstein Fellowship grant No. HF2-51541.001 awarded by the Space Telescope Science Institute (STScI), which is operated by the Association of Universities for Research in Astronomy, Inc., for NASA, under contract NAS5-26555. CL acknowledges support under DOE award DESC0010008 to Rutgers University and support from HST-GO-17474. MP is supported by NASA through a Einstein Fellowship grant No. HST-HF2-51583.001-A awarded by the Space Telescope Science Institute (STScI), which is operated by the Association of Universities for Research in Astronomy, Inc., for NASA, under contract NAS5-26555. E.E.H.\ is supported by a Gates Cambridge Scholarship (\#OPP1144). AA gratefully acknowledges support from AST2206195 (P.I. Narayan), to develop anomaly detection methods to identify lensed supernovae, and HST-GO-17128 (PI: R. Foley) to adapt BayeSN to model JWST and HST observations of type Ia supernovae. AA gratefully acknowledges support from NSF AST-2421845 and support from the Simons Foundation as part of the NSF-Simons SkAI Institute as a 2026 SkAI Graduate Fellow at UIUC. GN gratefully acknowledges NSF support from NSF CAREER grant AST-2239364, supported in-part by funding from Charles Simonyi, to model type Ia supernovae with ground- and space-based data, AST 2206195 to develop anomaly detection methods to identify lensed supernovae, and DOE support through the Department of Physics at the University of Illinois, Urbana Champaign (Grant No. 13771275) to deploy the lensed SN modeling pipeline from this work for the Vera C. Rubin Observatory. GN also gratefully acknowledges support from NSF AST-2421845 and support from the Simons Foundation as part of the NSF-Simons SkAI Institute, and NSF OAC1841625, OAC-1934752, OAC-2311355, AST-2432428 as part of the Scalable Cyberinfrastructure for Multi-messenger Astrophysics (SCIMMA) team.

\section*{Data Availability}

The SN H0pe data used in this analysis is available publicly in \citet{Pierel24}. The BayeSN-TD code is available at \url{https://github.com/bayesn/bayesn-td}.



\bibliographystyle{mnras}
\bibliography{refs} 



\appendix

\section{Phase-extended BayeSN model}
\label{appendix:extended_bayesn}

As described in Section \ref{extended_bayesn}, within this work we introduce a new, phase-extended version of the BayeSN model. The primary motivation for the development of this model was for analysis of glSNe Ia serendipitously discovered by JWST, such as SN H0pe and SN Encore, given the late phase coverage of these observations. Previous analysis of SN H0pe by \citet{Pierel24} applied a version of the model from \citet{W22} which incorporated $U$-band data and covered phases out to +50 days, applying linear extrapolation for phases later than this. This model was used given its combination of both phase and wavelength coverage, given the need for a model that also covered NIR wavelengths. We trained this new BayeSN model to further extend the phase coverage of the model, removing the need to rely on extrapolation and improving reliability at later phases, while maintaining wavelength coverage.

\begin{figure*}
\centering
\includegraphics[width = \linewidth]{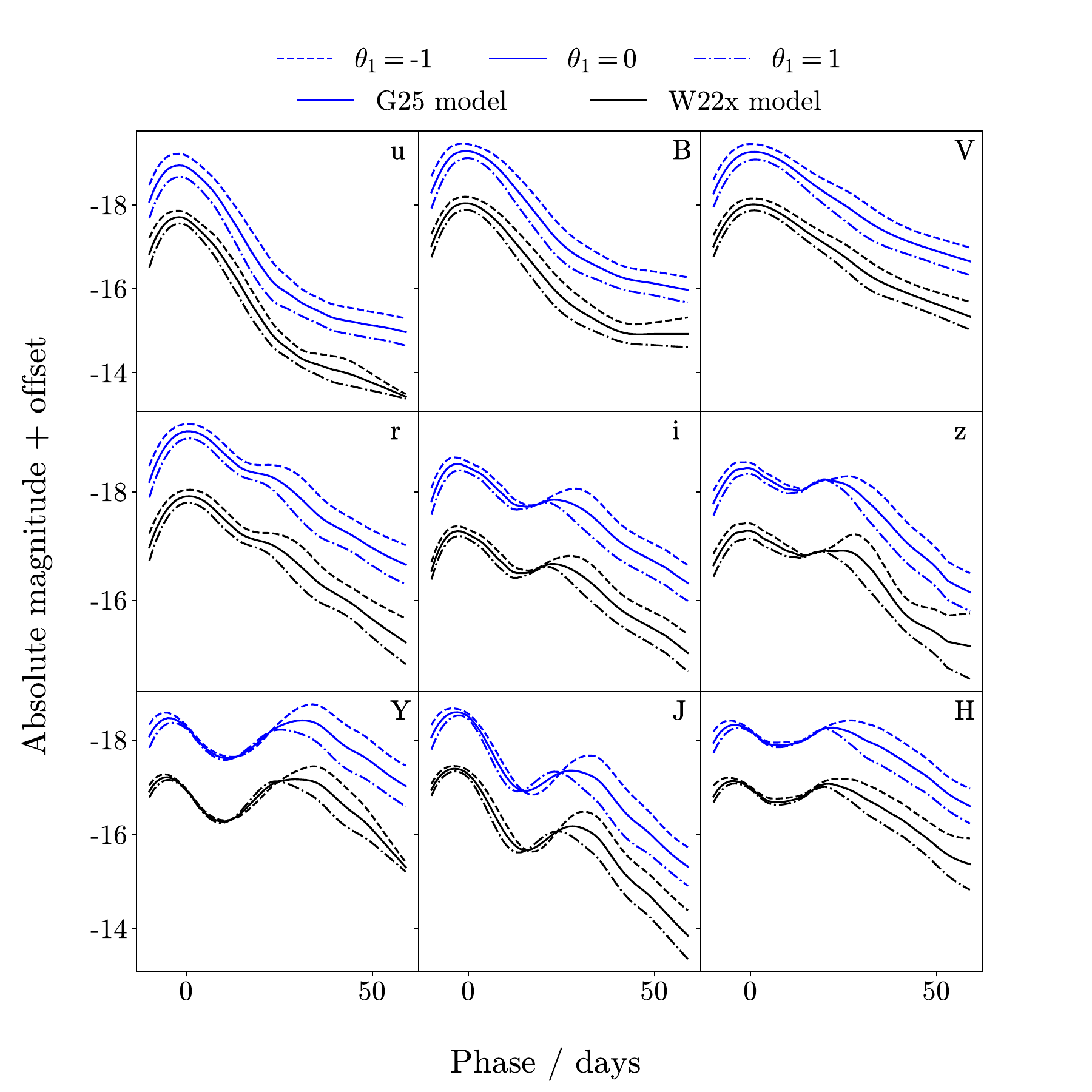}
\caption{Model light curves for the new, phase-extended BayeSN model presented in this work (G25 model) shown in blue, along with model light curves from the extended W22 model from \citet{W22} applied in \citet{Pierel24} (W22x model) shown in black. These were simulated with $A_V=\epsilon(t,\lambda_r)=\delta M=0$, and showcase the influence of the functional principal component score $\theta_1$ on the BayeSN model SN Ia light curve across numerous bands. The light curves cover the phase range over which data is available for SN H0pe, to allow for comparison with the extended W22x model. The W22x lines are arbitrarily offset for clarity.}
\label{new_model_lcplots}
\end{figure*}

Fig. \ref{new_model_lcplots} demonstrates the behaviour of this new BayeSN model; this figure shows example light curves evaluated for $\theta_1=-1,0,1$ with $A_V=\epsilon(t,\lambda_r)=\delta M=0$, to demonstrate a typical SN light curve for this model and how it is influenced by the functional principal component score $\theta_1$. To showcase how this model has improved reliability at later phases relative to the phase-extended model used in \citet{Pierel24}, this figure also shows the same but for the previous model. This plot shows model SN light curves for a phase range of -10 days to +60 days to allow for direct comparison, with +60 days covering the approximate phase range of SN H0pe. It is noticeable that the newer model demonstrates much more physical behaviour at later times, more in line with the expected linear decline in mag space resulting from a radioactive decay-driven light curve. While this new BayeSN model was trained on data out to phases as late as +85 days, available training data becomes increasingly scarce at these later phases and the model likely becomes increasingly less reliable. Within this work, we demonstrate the overall reliability of this phase-extended BayeSN model by successfully applying it for time-delay estimation of simulations which are based on an alternative SED model, SALT. For future analyses which focus on later-phase observations of SNe Ia, an increased training sample at later phases would be valuable to improve the reliability of SED models in general.

\section{Lens Model Weights}
\label{lens_weights}

As discussed in Section \ref{H0_constraints}, each individual lens model is weighted according to the consistency between the magnifications and time delays predicted by each modes and estimated using SN H0pe. The specific weights for each model across our analysis variants are presented in Table \ref{lens_weights_table}, with the lens models themselves listed in Table \ref{tab:lens_models}. Full discussion of the lens models and the weighting scheme are presented in \citet{Pascale25}, but we include some details here for completeness.

We follow \citet{Pascale25} by denoting the observed SN light curve as LC, which leads to a set of observables $\mathcal{O}$. These observables depend on the analysis variant being considered; in our fiducial case, $\mathcal{O}: \{\Delta t_{BA}, \Delta t_{BC}, \beta_A, \beta_B, \beta_C\}$, while for the TD-only case we consider only the time delays. Our goal in this analysis is to determine the posterior distribution of $H_0$, $P(H_0|\text{LC})$, marginalised over each lens model $M_l$. Each lens model makes predictions of $\mathcal{O}$ assuming a fiducial $H_0$ value. For each individual model, the fiducial $H_0$ value is rescaled to match the model predictions to the inferred values from the LC. Combining the constraints from each lens model, the posterior on $H_0$ is:

\begin{equation}
P(H_0 \mid \mathrm{LC})
\propto
P(H_0)\sum_{i=1}^{N}
\int
P(\mathcal{O}\mid M_i; H_0)\,
P(\mathcal{O}\mid \mathrm{LC})\,
d\mathcal{O}_1 \cdots d\mathcal{O}_n \, 
\end{equation}

with weights determined by assessing $\int P(\mathcal{O}\mid M_i; H_0)\, P(\mathcal{O}\mid \mathrm{LC})\,P(H_0)\, dH_0\, d\mathcal{O}_1 \cdots d\mathcal{O}_n$. This approach ensures that lens models with predictions most consistent with $\mathcal{O}$ hold the greatest weight in the overall $H_0$ inference.

\begin{table*}
\centering
\caption{Lens models weights for each of our analysis variants. Note that Model 6 has a weight of 0 because the PDF generally falls outside the range of the $H_0$ prior.}
\begin{tabular}{lccccccc}
\hline
Analysis & Model 1 & Model 2 & Model 3 & Model 4 & Model 5 & Model 6 & Model 7 \\
\hline
Weighted Phot-only  & $1\times10^{-5}$ & $4\times10^{-3}$ & $2\times10^{-3}$ & 0.805 & 0.194 & 0 & $2\times10^{-5}$ \\
Weighted Phot+spec  & $4\times10^{-5}$ & $5\times10^{-5}$ & $7\times10^{-5}$ & 0.542 & 0.458 & 0 & $8\times10^{-5}$ \\
TD-only Phot-only   & 0.282  & 0.086  & 0.317  & 0.067  & 0.248  & 0      & 0.0002 \\
TD-only Phot+spec   & 0.282  & 0.086  & 0.317  & 0.067  & 0.248  & 0      & 0.0002 \\
\hline
\end{tabular}
\label{lens_weights_table}
\end{table*}

\begin{table*}
\centering
\caption{SN H0pe lens models used in \citet{Pascale25} and this work.}
\label{tab:lens_models}
\begin{tabular}{cll}
\hline
Model Number & Lens Model        & Citations \\ \hline
1        & GLAFIC                 & \citep{Oguri10, Oguri21} \\
2        & Zitrin-analytic        & \citep{Zitrin15} \\
3        & LENSTOOL               & \citep{Kneib11} \\
4        & MARS                   & \citep{ChaJee22} \\
5        & Chen    & \citep{Chen20} \\
6        & WSLAP+                 & \citep{Diego05} \\
7        & Zitrin-LTM             & \citep{Zitrin09} \\
\hline
\end{tabular}
\end{table*}


\bsp	
\label{lastpage}
\end{document}